\documentclass[sigconf]{acmart}
\usepackage{colortbl}
\usepackage{multirow}
\usepackage{makecell}
\usepackage{subfigure}
\usepackage{enumitem}

\AtBeginDocument{%
  }

\setcopyright{acmlicensed}
\copyrightyear{2018}
\acmYear{2018}
\acmDOI{XXXXXXX.XXXXXXX}
\acmConference[Conference acronym 'XX]{Make sure to enter the correct
  conference title from your rights confirmation email}{June 03--05,
  2018}{Woodstock, NY}
\acmISBN{978-1-4503-XXXX-X/2018/06}

\begin{document}

\title{RaSeRec: Retrieval-Augmented Sequential Recommendation}


\author{Xinping Zhao$^1$, Baotian Hu$^{1*}$, Yan Zhong$^2$, Shouzheng Huang$^1$, Zihao Zheng$^1$, Meng Wang$^3$, \\ Haofen Wang$^{3*}$, and Min Zhang$^1$}

\affiliation{
\institution{$^1$Harbin Institute of Technology (Shenzhen) \city{Shenzhen} \country{China}}
}
\affiliation{
\institution{$^2$Peking University \city{Beijing} \country{China}, $^3$Tongji University \city{Shanghai} \country{China}}
}
\email{zhaoxinping@stu.hit.edu.cn, mengwangtj@tongji.edu.cn}
\email{carter.whfcarter@gmail.com,{hubaotian, zhangmin2021}@hit.edu.cn}

\thanks{$^*$Baotian Hu and Haofen Wang are the corresponding authors}

\renewcommand{\shortauthors}{Zhao et al.}

\begin{abstract}
Although prevailing supervised and self-supervised learning augmented sequential recommendation (SeRec) models have achieved improved performance with powerful neural network architectures, we argue that they still suffer from two limitations: 
\textbf{(1) Preference Drift}, where models trained on past data can hardly accommodate evolving user preference; 
and \textbf{(2) Implicit Memory}, where head patterns dominate parametric learning, making it harder to recall long tails. 
In this work, we explore retrieval augmentation in SeRec, to address these limitations.
Specifically, we propose a {\underline{R}}etrieval-{\underline{A}}ugmented {\underline{Se}}quential {\underline{Rec}}ommendation framework, named \textbf{RaSeRec}, the main idea of which is to maintain a dynamic memory bank to accommodate preference drifts and retrieve relevant memories to augment user modeling explicitly.
It consists of two stages: {(i) collaborative-based pre-training}, which learns to recommend and retrieve; {(ii) retrieval-augmented fine-tuning}, which learns to leverage retrieved memories.
Extensive experiments on three datasets fully demonstrate the superiority and effectiveness of RaSeRec. The implementation code is available at \url{https://github.com/HITsz-TMG/RaSeRec}.
\end{abstract}




\maketitle

\section{Introduction}
\label{sec:intro}
\begin{figure}[t]
    \centering
    \includegraphics[width=1.0\linewidth]{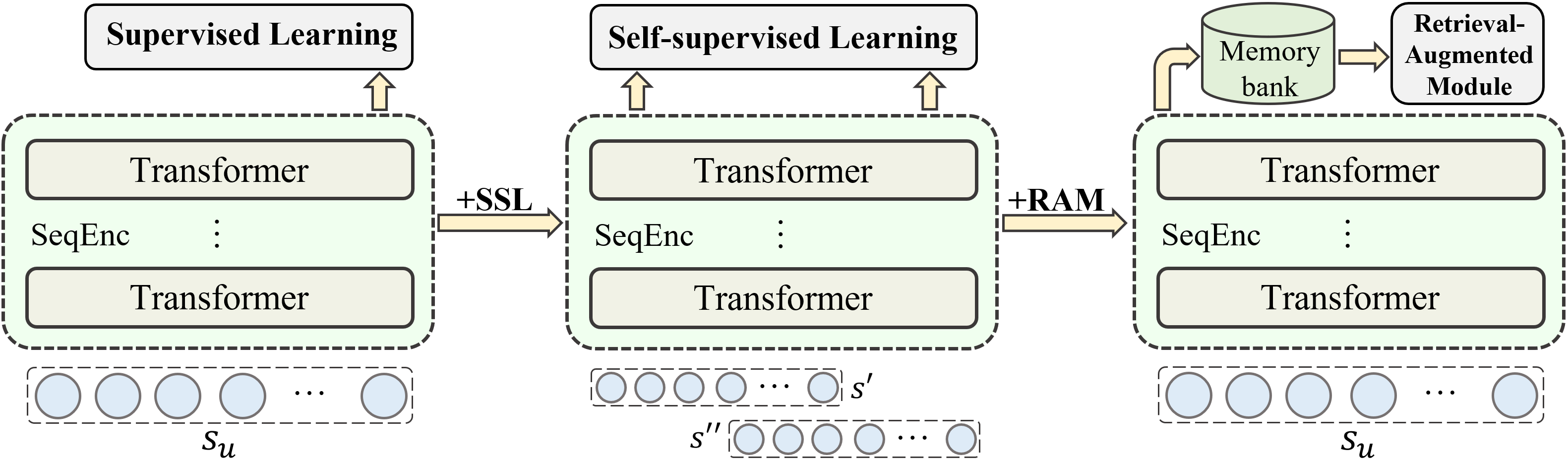}
    \caption{Comparison between Vanilla, SSL-Augmented, and Retrieval-Augmented SeRec paradigms, where their system working flows are illustrated from left to right, respectively. $s_u$ denotes the user sequence, while $s_u^\prime, s_u^{\prime\prime}$ represent two augmented views. SeqEnc denotes the sequence encoder, referring to \S\ref{sec:trans_sr} for more technical details.}
    \label{fig:comparison_paradigm}
    \vspace{-0.25cm}
\end{figure}

Sequential Recommendation (SeRec) aims to predict the next items that users would like to adopt, by calculating the matching probabilities between their historical interactions and the candidate items \cite{DBLP:conf/icdm/KangM18,DBLP:conf/icws/Zhao0SZH24,DBLP:journals/corr/HidasiKBT15}. 
As such, it is essential to learn high-quality user representations from the user's historical interactions for better recommendation performance \cite{DBLP:conf/icde/XieSLWGZDC22,DBLP:conf/wsdm/QiuHYW22,DBLP:journals/corr/abs-2108-06479}. 
However, with enormous parameters to be optimized, existing SeRec models commonly suffer from the data sparsity issue, 
\textit{i.e.,} users only interacted with a tiny fraction of the total items \cite{isinkaye2015recommendation}, making it difficult to learn high-quality user representations.
Recently, there has been a surge of research interest in leveraging self-supervised learning (SSL) within the realm of SeRec to alleviate the data sparsity issue caused by sparse supervised signals \cite{DBLP:conf/icde/XieSLWGZDC22,DBLP:conf/wsdm/QiuHYW22,DBLP:journals/corr/abs-2108-06479,DBLP:conf/icde/HaoZFQ0ZS024,DBLP:conf/aaai/DangYGJ0XSL23}. 
Generally, it involves two procedures: \textbf{(1) data augmentation}, which generates multiple views for each user sequence, and \textbf{(2) self-supervised learning}, which maximizes the agreement between different views of the same (or similar) user sequence.
In this way, SSL-Augmented models endow their model backbones with additional parametric knowledge to generate more high-quality user representations than those without SSL \cite{DBLP:conf/sigir/WuWF0CLX21}.
Despite their effectiveness, we argue that current SSL-Augmented ones still suffer from some severe limitations:
\begin{itemize}[leftmargin=*]
    \item \textbf{Preference Drift.} In real-world recommender systems, user preference frequently drifts as time goes on \cite{DBLP:journals/tkde/ChenLYY22,DBLP:journals/tcss/LoLCL18} due to many reasons, \textit{e.g.,} life changes \cite{DBLP:conf/recsys/AghdamHMB15}. 
    As such, models trained on past user preferences can hardly handle preference drifts and may recommend undesirable items to users, which leads to sub-optimal performance. 
    
    \item \textbf{Implicit Memory.} Existing methods encode user sequential patterns into implicit memory (model parameters), where the distribution of observed patterns usually obeys a power law \cite{DBLP:journals/siamrev/ClausetSN09,DBLP:journals/jasis/Milojevic10a}. As such, long-tailed patterns that lack supervised signals may be overwhelmed by head ones during training, making it hard to recall long-tailed ones during inference.
\end{itemize} 
In this work, we explore Retrieval-Augmented Generation (RAG) \cite{DBLP:journals/corr/abs-2312-10997} in SeRec, to address the above limitations. Though being well-studied in Large Language Models (LLMs) \cite{DBLP:conf/nips/LewisPPPKGKLYR020,DBLP:conf/iclr/IzacardG21,DBLP:journals/jmlr/IzacardLLHPSDJRG23,zhao2024funnelrag,DBLP:conf/iclr/AsaiWWSH24}, RAG is rarely explored in SeRec. 
The main idea of RAG is to retrieve relevant documents from an external knowledge base to facilitate LLMs' generation, especially for knowledge-intensive tasks \cite{DBLP:conf/nips/LewisPPPKGKLYR020}. 
For example, RAG \cite{DBLP:conf/nips/LewisPPPKGKLYR020} uses the input sequence as the query to retrieve text documents from Wikipedia and then adopts them as additional context to generate the target sequence; Self-RAG \cite{DBLP:conf/iclr/AsaiWWSH24} learns to retrieve passages on demand adaptively, and interleaves generation and retrieval with reflection tokens by unifying them as the next token prediction. 
Compared to SSL (similar to pre-training), RAG allows us to accommodate {preference drift} rapidly and recall {long tails} via looking up an external memory bank directly.
Here, we wish to bring the RAG's superiority into SeRec to enhance user representation learning, which differs from NLP tasks since they model \textit{language semantics} while recommender systems model \textit{collaborative semantics} \cite{DBLP:conf/icde/ZhengHLCZCW24}. 
To address the above limitations of SSL-Augmented SeRec models, we design a retrieval-augmented mechanism to improve the quality of user representations. 
Specifically, it consists of two key components: \textbf{(1) memory retrieval}, which recalls collaborative memories for the input user, and \textbf{(2) representation augmentation}, which leverages retrieved memories to augment user representation. 
On the one hand, \textit{memory retrieval} adapts well to dynamic preference drifts with a real-time update memory bank. 
On the other hand, \textit{representation augmentation} explicitly leverages retrieved memories, rendering it easier to recall long tails. 
Figure \ref{fig:comparison_paradigm} illustrates the difference between the three SeRec paradigms and the continuation between them.
The vanilla SeRec is trained on supervised signals to predict the next item based on the input user sequence \cite{DBLP:journals/tois/FangZSG20}.
The SSL-Augmented SeRec further enriches the parametric knowledge of the model backbone with SSL signals \cite{DBLP:journals/tkde/YuYXCLH24}. 
Standing on their shoulders, the retrieval-augmented SeRec handles dynamic preference drifts and relieves the heavy reliance on implicit memory by \textit{memory retrieval} and \textit{representation augmentation}. Note that the retrieval-augmented SeRec is designed to further perfect existence, rather than completely replacing them.
To this end, we propose a novel \underline{R}etrieval-\underline{A}ugmented \underline{Se}quential \underline{Rec}ommendation paradigm, named \textbf{RaSeRec}, which learns to refine user representations with retrieved memories. 
Specifically, RaSeRec’s training process consists of two key stages: (i) collaborative-based pre-training and (ii) retrieval-augmented fine-tuning. 
In the \textbf{collaborative-based pre-training} stage, the model is trained with two objectives built upon collaborative signals: {(1) recommendation learning}, which learns to predict the next item based on input user sequence, 
and {(2) retrieval training}, which learns to retrieve memories with the same (or similar) preference as the input user sequence. 
During the \textbf{retrieval-augmented fine-tuning} stage, we treat <user sequence, target item> pairs as the \textit{reference set} except the current input user sequence. 
We then use the pre-trained model to encode the reference set into a memory bank. 
Given a user sequence, our model first retrieves similar collaborative memories from the memory bank. 
Then, a Retrieval-Augmented Module (RAM) is built upon the pre-trained encoder, which learns to augment the representation of the input user with retrieved memories.  
An intuitive explanation of our idea is that it can be seen as an open-book exam. Specifically, with retrieval augmentation, RaSeRec does not have to memorize all sequential patterns whereas it learns to use retrieved ones (cheat sheets) from the memory bank.
Note that RaSeRec is model-agnostic and can be applied to any ID-based SeRec backbones. Here, we adopt SASRec \cite{DBLP:conf/icdm/KangM18} as the backbone since it is simple yet effective. 
We also implement it on different backbones (see \S\ref{sec:impro_base}).
In summary, our contributions are: 
\begin{itemize}[leftmargin=*]
    \item We unveil the major issues existing models suffer from and shed light on the potential of RAG in SeRec. 
    \item We propose a new SeRec paradigm, RaSeRec, which explicitly retrieves collaborative memories and then learns to use them for better user representation modeling.
    \item Extensive experiments on three benchmark datasets fully demonstrate that RaSeRec improves overall performance, enhances long-tailed recommendation, and alleviates preference drift. 
\end{itemize}

\section{Preliminaries}
\label{sec:preliminary}
\subsection{Problem Formulation}
In this section, we first introduce the symbols and then formalize the problem of SeRec. 
Let $\mathcal{U}$ and $\mathcal{V}$ denote the set of users and items, respectively, where $|\mathcal{U}|$ and $|\mathcal{V}|$ denote the number of users and items. 
The historically interacted items of a user are sorted in chronological order: $s_u = [v_1^{(u)}, v_2^{(u)}, ...,v_{|s_u|}^{(u)}]$, where
$v_t^{(u)} \in \mathcal{V}, 1 \leq t \leq |s_u|$ denotes the item interacted by user $u \in \mathcal{U}$ at the time step $t$ and $|s_u|$ denotes the length of interaction sequence of user $u$. 
Besides, $s_{u,t} = [v_1^{(u)}, v_2^{(u)}, ...,v_{t}^{(u)}]$ is a subsequence, where items are interacted by user $u$ before the time step $t+1$.
The goal of sequential recommendation is to predict the next item $v_{|s_u|+1}^{(u)}$ given the interaction sequence $s_u$, which can be formulated as follows:
\begin{equation}
v_* =  \arg \underset{v_i\in \mathcal{V}}\max {P\left( v_{|s_u|+1}^{(u)} = v_i | s_u\right)},
\end{equation}
where $v_*$ denotes the predicted item with the largest matching probability. 
In what follows, we employ bold lowercase and uppercase symbols to represent vectors and matrices, respectively.

\subsection{Transformer for SeRec}
\label{sec:trans_sr}
In this section, we introduce how to model users' historical interactions to get their representations. Specifically, we adopt SASRec \cite{DBLP:conf/icdm/KangM18} as the model backbone, whose encoding module
is based on the Transformer encoder \cite{DBLP:conf/nips/VaswaniSPUJGKP17}. 
To leverage the transformer's strong encoding ability, we first convert items to embeddings. Then, we apply the Transformer encoder to generate the user representation. 
\subsubsection{Embedding Layer} Formally, an embedding table ${\mathbf V} \in \mathbb{R}^{|\mathcal{V}| \times d}$ is created upon the whole item set $\mathcal{V}$ to project each item (one-hot representation) into a low-dimensional dense vector for better sequential modeling, where $d$ is the embedding dimension. 
Additionally, to perceive the time order of the sequence, a learnable position encoding matrix $\mathbf{P} \in \mathbb{R}^{T \times d}$ is constructed, where $T$ is the maximum sequence length. Owing to the constraint of the maximum sequence length $T$, when inferring user $u$ representation at time step $t+1$, we truncate the user sequence $s_{u,t}$ to the last $T$ items if $t > T$, otherwise left unchanged:
\begin{equation}
    s_{u,t} = [v_{t-T+1}, v_{t-T+2}, ..., v_{t}],
\end{equation}
where we omit the superscript ${(u)}$ for brevity. Subsequently, the item embedding and position encoding are added together to make up the input vector:
\begin{equation}
    {\mathbf h}_i^0 = {\mathbf v}_i+{\mathbf p}_i,
\end{equation}
where ${\mathbf h}_i^0$ is the input vector at position $i$. All input vectors collectively form the hidden representation of user sequence $s_{u,t}$ as $\mathbf{H}^0 = [\mathbf{h}_{t-T+1}^0, ..., \mathbf{h}_t^0]$. 
\subsubsection{Transformer Encoder} A Transformer encoder is composed of a multi-head self-attention module and a position-wise feed-forward Network. 
More technical details about these two modules can be found in \cite{DBLP:conf/icdm/KangM18} and \cite{DBLP:conf/nips/VaswaniSPUJGKP17}. We stack multiple blocks to capture more complex sequential patterns for high-quality user representation. 
Specifically, an ${L}$-layer Transformer encoder (unidirectional attention) is applied to refine the hidden representation of each item in ${\mathbf H}^0$, which can be formulated as follows:
\begin{equation}
    {\mathbf H}^L = \rm{Trm}({\mathbf H}^0),
\end{equation}
where $\rm{Trm}(\cdot)$ denotes the $L$-layer Transformer encoder, ${\mathbf H}^L = [\mathbf{h}_{t-T+1}^L, ..., \mathbf{h}_t^L]$ is the refined hidden representations of the sequence. 
The last refined hidden representation $\mathbf{h}_t^L$ is selected as the representative of the user sequence. Here, we omit the superscript $L$ and subscript $t$ for brevity, \textit{i.e.,} $\mathbf{h}$. 
In addition, we use $\mathrm{SeqEnc(\cdot)}$ to denote the sequence encoder, that is $\mathbf{h} = \mathrm{SeqEnc}(s_{u,t})$ for simplicity.
\begin{figure}[t]
    \centering
    \includegraphics[width=1.0\linewidth]{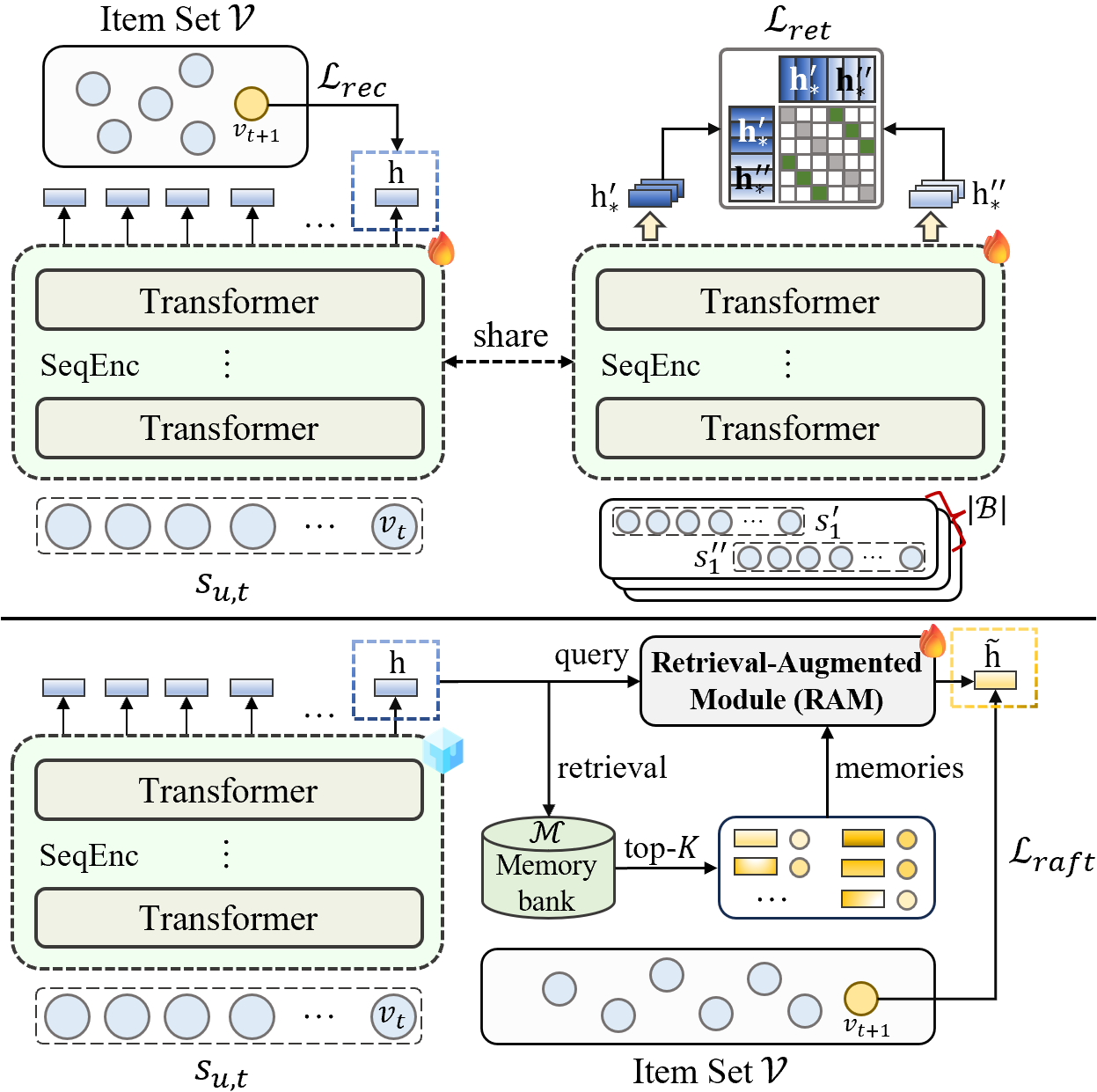}
    \caption{The overall system framework of the proposed RaSeRec. The upper layer illustrates the workflow of collaborative-based pre-training while the bottom layer shows the workflow of retrieval-augmented fine-tuning.}
    \label{fig:raserec_framework}
    \vspace{-0.2cm}
\end{figure} 
\section{Methodology}
We propose the Retrieval-Augmented Sequential Recommendation (RaSeRec) paradigm, which endows SeRec to accommodate preference drift and recall long-tailed patterns, as shown in Figure \ref{fig:raserec_framework}. 
Specifically, it consists of a two-stage training strategy: \textbf{(1) collaborative-based pre-training}, which learns to recommend and retrieve based on collaborative signals (\S \ref{sec:colla_pt}); 
\textbf{(2) retrieval-augmented fine-tuning}, which learns to leverage retrieved memories to augment user representation (\S \ref{sec:retri_ft}).
Lastly, we introduce the inference process and analyze the complexity of RaSeRec (\S \ref{sec:infer_complex}). 
The notation table can be found in Appendix \ref{app:notation}. 
\subsection{Collaborative-based Pre-training}
\label{sec:colla_pt}
In this section, we pre-train the model backbone $\mathrm{SeqEnc}(\cdot)$ with two learning objections: \textbf{(1) recommendation learning}, which endows the backbone with the ability to generate the next item based on the input user sequence, 
and \textbf{\textbf{(2) retrieval training}}, which endows the backbone with the ability to retrieve memories with the same (or similar) preference as the input user sequence. 
\subsubsection{Recommendation Learning} After computing the user representation $\mathbf{h} \in \mathbb{R}^d$ (\S\ref{sec:trans_sr}), a prediction layer is built upon it to predict how likely user $u$ would adopt item $v_i$ at time step $t+1$. 
A simple yet effective solution is to calculate the inner product, \textit{i.e.,} $\mathbf{h}^T\textbf{v}_i$. After that, we employ the next item prediction over the whole item set $\mathcal{V}$ as the recommendation learning objective. 
To be specific, we compute the negative log-likelihood with the softmax as the recommendation learning objective:
\begin{equation}
\label{equ:rec}
     \mathcal{L}_{rec} = -\log \Big( e^{\mathbf{h}^T \mathbf{v}_{t+1}} / \sum_{v_i \in \mathcal{V}} e^{\mathbf{h}^T \mathbf{v}_{i}}\Big),
\end{equation}
where $\mathbf{h}, \mathbf{v}_{t+1}, \mathbf{v}_{i}$ denote the last refined hidden representation $\mathbf{h}_t^{L}$, the embedding vector of the next item $v_{t+1}$, and the embedding vector of item $v_i \in \mathcal{V}$, respectively. 
We train the model to predict the next item auto-regressively similar to the language modeling's objective \cite{Radford2018ImprovingLU}. 
\subsubsection{Retrieval Training} To retrieve memories that share the same (or similar) preference as the input user sequence, a retrieval training objective is developed to maximize the agreement of positive pairs and minimize that of negative pairs. 
The positive pair consists of two user sequences that share the same (or similar) preference. In SeRec, the goal is to predict the next item. Therefore, if the next item of two user sequences is the same, these two sequences can be thought to have similar collaborative semantics \cite{DBLP:conf/wsdm/QiuHYW22}. 
Particularly, the positive pair can be defined as $s_{u^\prime, t^i} = [v_{t^i-T+1}, ..., v_{t^i}],$ and $s_{u^{\prime\prime}, t^j} = [v_{t^j-T+1}, ..., v_{t^j}]$, where $v_{t^i+1}$ and $v_{t^j+1}$ are the same item. For brevity, we mark $s_{u^\prime, t^i}, s_{u^{\prime\prime}, t^j}$ as $s^{\prime}, s^{\prime\prime}$, respectively. The user representation of them can be defined as:
\begin{equation}
\label{equ:pos1}
    \mathbf{h}^{\prime} = \mathrm{SeqEnc(s^{\prime})}, \quad \mathbf{h}^{\prime\prime} = \mathrm{SeqEnc(s^{\prime\prime})}.
\end{equation}
In retrieval training settings, negative samples usually need to be selected from a large pool \cite{DBLP:conf/emnlp/KarpukhinOMLWEC20}. 
To efficiently construct negatives, we treat the other samples in the same training batch as negatives. Assume that we have a training batch $\mathcal{B}$ that consists of $|\mathcal{B}|$ positive pairs:
\begin{equation}
\label{equ:pos2}
    \mathcal{B} = \{\langle\mathbf{h}^{\prime}_1, \mathbf{h}^{\prime\prime}_1\rangle, \langle\mathbf{h}^{\prime}_2, \mathbf{h}^{\prime\prime}_2\rangle, ..., \langle\mathbf{h}^{\prime}_{|\mathcal{B}|}, \mathbf{h}^{\prime\prime}_{|\mathcal{B}|}\rangle\}.
\end{equation}
As such, for each positive pair, there are $2(|\mathcal{B}|-1)$ negatives forming the negative set $\mathcal{S}^-$. For example, for the positive pair $\langle\mathbf{h}^{\prime}_1, \mathbf{h}^{\prime\prime}_1\rangle$, its corresponding negative set is $\mathcal{S}^-_1=\{\mathbf{h}^{\prime}_2, \mathbf{h}^{\prime\prime}_2, ..., \mathbf{h}^{\prime}_{|\mathcal{B}|}, \mathbf{h}^{\prime\prime}_{|\mathcal{B}|}\}$. 
Having established the positive and negative pairs, we adopt InfoNCE \cite{DBLP:journals/jmlr/GutmannH10} as the loss function, which can be defined as follows:
\begin{equation}
\label{equ:ret}
\begin{split}
    \mathcal{L}_{ret} = &- \Big( {
    \log  \frac{e^{\mathrm{s}(\mathbf{h}^{\prime}_i, \mathbf{h}^{\prime\prime}_i)/\tau}}{
    e^{\mathrm{s}(\mathbf{h}^{\prime}_i, \mathbf{h}^{\prime\prime}_i)/\tau} + 
    \sum_{s^{-} \in 
    \mathcal{S}^-_{i}} e^{\mathrm{s}(\mathbf{h}^{\prime}_i, \mathbf{h}^{-})/\tau}}} \\
    &+ {
    \log  \frac{e^{\mathrm{s}(\mathbf{h}^{\prime\prime}_i, \mathbf{h}^{\prime}_i)/\tau}}{
    e^{\mathrm{s}(\mathbf{h}^{\prime\prime}_i, \mathbf{h}^{\prime}_i)/\tau} + 
    \sum_{s^{-} \in 
    \mathcal{S}^-_{i}} e^{\mathrm{s}(\mathbf{h}^{\prime\prime}_i, \mathbf{h}^{-})/\tau}}} \Big),
\end{split}
\end{equation}
where we take the $i$-th positive pair as an example; $\mathcal{S}^-_{i}$ denotes the negative set; $s(\cdot)$ measures the similarity between two vectors, which is implemented as the inner product; $\tau$ is the temperature, which is used to control the smoothness of softmax. We pre-train the model backbone with the training objectives of $\mathcal{L}_{rec}$ as well as $\mathcal{L}_{ret}$, simultaneously.
\begin{figure}[t]
    \centering
    \includegraphics[width=1.0\linewidth]{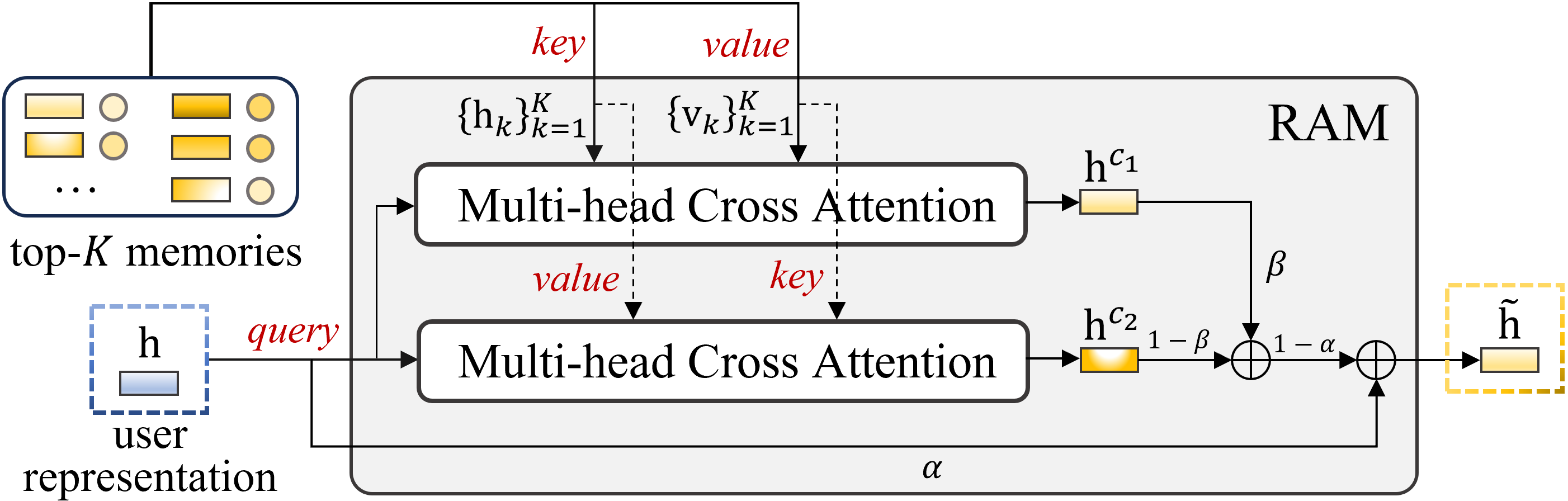}
    \caption{The architecture of Retrieval-Augmented Module, where $\tilde{\mathbf{h}}$ is the augmented user representation.}
    \label{fig:ram}
    \vspace{-0.25cm}
\end{figure}
\subsection{Retrieval-Augmented Fine-tuning}
\label{sec:retri_ft}
After pre-training, the model backbone has been endowed with the ability to recommend the next item and retrieve relevant memories. 
However, it still suffers from two limitations: 
\textbf{(i)} It can hardly accommodate dynamic \textit{preference drifts} as it is trained on past user preferences, and 
\textbf{(ii)} It may fail to recall long-tailed patterns from \textit{implicit memory} due to skewed data distribution. 
Given the above issues, we propose Retrieval-Augmented Fine-Tuning, RAFT, making the model quicker to adapt to preference drifts and easier to recall long-tailed patterns.
It mainly consists of two components: \textbf{(1) memory retrieval}, which recalls useful memories from the memory bank, 
and \textbf{(2) representation augmentation}, which leverages the retrieved memories to augment user representation. 
\subsubsection{Memory Retrieval} Collaborative <user sequence $s_{u,t}$, target item $v_{t+1}^{(u)}$> pairs explicitly reveal the sequential patterns, where pairs sharing the same (or similar) preference as the input user sequence can act as important references. 
Inspired by the above idea, we construct a reference set $\mathcal{R}$ by auto-regressively enumerating all <$s_{u,t}$, $v_{t+1}^{(u)}$> pairs from the training data except the current input user sequence. 
We then employ the sequence encoder $\mathrm{SeqEnc}(\cdot)$ to encode all pairs in $\mathcal{R}$ so that we can get a memory bank $\mathcal{M}$, where each entry in $\mathcal{M}$ consists of a user representation and the corresponding target item embedding.
Given the input representation $\mathbf{h}$, RaSeRec retrieves top-$K$ similar user representations $\{ {\mathbf{h}}_{k}\}_{k=1}^K$ as well as their corresponding target item embedding $\{ {\mathbf{v}}_{k}\}_{k=1}^K$ from the memory bank $\mathcal{M}$,
where we employ the Faiss library \cite{DBLP:journals/tbd/JohnsonDJ21} to speed up the retrieval process.
\subsubsection{Representation Augmentation} With the retrieved memories, we design a Retrieval-Augmented Module (RAM), which learns to leverage the retrieved memories to augment the current user representation. 
Its overall architecture is illustrated in Figure \ref{fig:ram}. Inspired by \cite{DBLP:conf/cvpr/XieSXZZZ23}, RAM employs dual-channel multi-head cross attention (MHCA) which takes $\mathbf{h}$ as \textit{query}, $\{ {\mathbf{h}}_{k}\}_{k=1}^K$ (or $\{ {\mathbf{v}}_{k}\}_{k=1}^K$) as \textit{key} and $\{ {\mathbf{v}}_{k}\}_{k=1}^K$ (or $\{ {\mathbf{h}}_{k}\}_{k=1}^K$) as \textit{value} to get the augmented user representation. 
In the first channel, RAM learns to weight and aggregate $\{ {\mathbf{v}}_{k}\}_{k=1}^K$ via modeling the relationship between $\mathbf{h}$ and $\{ {\mathbf{h}}_{k}\}_{k=1}^K$:
\begin{equation}
\label{eq:first_channel}
    \mathbf{h}^{c_1} = \mathrm{MHCA}(\mathbf{h}, \{ {\mathbf{h}}_{k}\}_{k=1}^K, \{ {\mathbf{v}}_{k}\}_{k=1}^K),
\end{equation}
where $\mathrm{MHCA}(\cdot)$ is the multi-head cross attention module; $\mathbf{h}^{c_1}$ is the first augmented representation. 
Analogously, RAM compute the second augmented representation $\mathbf{h}^{c_2}$ through aggregating $\{ {\mathbf{h}}_{k}\}_{k=1}^K$:
\begin{equation}
\label{eq:second_channel}
    \mathbf{h}^{c_2} = \mathrm{MHCA}(\mathbf{h}, \{ {\mathbf{v}}_{k}\}_{k=1}^K, \{ {\mathbf{h}}_{k}\}_{k=1}^K).
\end{equation}
Lastly, RAM weights them together and obtains the final retrieval-augmented user representation: 
\begin{equation}
\label{eq:fusion}
    \tilde{\mathbf{h}} = \alpha \mathbf{h} + (1-\alpha)(\beta\mathbf{h}^{c_1}+(1-\beta)\mathbf{h}^{c_2}),
\end{equation}
where $0 \leq \alpha, \beta \leq 1$ control the strength of each representation. A possible extension is to learn $\alpha$ and $\beta$, \textit{e.g.,} designing attention mechanism \cite{DBLP:conf/aaai/WuT0WXT19}. 
However, it is not the focus of this work, and we leave it for future exploration. We further fine-tune the model with RAFT to improve recommendations with the retrieval-augmented representation $\tilde{\mathbf{h}}$. 
Specifically, we freeze the model backbone $\mathrm{SeqEnc}(\cdot)$ and only the RAM
is being updated during fine-tuning. Formally, we adopt the same objective as Equ (\ref{equ:rec}), which learns to leverage the retrieved memories for better recommendation:
\begin{equation}
\label{eq:raft}
    \mathcal{L}_{raft} = -\log \Big( e^{\tilde{\mathbf{h}}^T \mathbf{v}_{t+1}} / \sum_{v_i \in \mathcal{V}} e^{\tilde{\mathbf{h}}^T \mathbf{v}_{i}}\Big).
\end{equation}
In this way, RaSeRec does not have to memorize all long-tailed patterns but learns to leverage retrieved explicit memories to accommodate preference drift.

\begin{table*}[!t]
\centering
\footnotesize
\tabcolsep=0.2cm
\renewcommand\arraystretch{1.1}
\caption{The space and time complexity comparison between the Vanilla and SSL/Retrieval-Augmented SeRec.}
\begin{tabular}{c|cc|cccc}
\toprule
    \multirow{2}{*}{\textbf{Paradigms}} &
    \multicolumn{2}{c}{\textbf{Space}} &
    \multicolumn{4}{|c}{\textbf{Time}} 
    \\
    \cline{2-7}
    &  \multicolumn{1}{c}{\centering Backbone} & \multicolumn{1}{c}{\centering RAM} & \multicolumn{1}{|c}{\centering Encode} & \multicolumn{1}{c}{\centering Retrieve} & \multicolumn{1}{c}{\centering Augment} & \multicolumn{1}{c}{\centering Recommend}\\ 
    \hline
    \hline
    \multirow{1}{*} {\centering {Vanilla SeRec}} & \multicolumn{1}{c}{\centering $O(|\theta|)$} & \multicolumn{1}{c|}{\centering -}  & \multicolumn{1}{c}{\centering $O(L(Td^2+T^2d))$} & \multicolumn{1}{c}{\centering -} & \multicolumn{1}{c}{\centering -} & \multicolumn{1}{c}{\centering $O(|\mathcal{V}|d)$}\\
     \multirow{1}{*} {\centering {SSL-Augmented SeRec}} & \multicolumn{1}{c}{\centering $O(|\theta|)$} & \multicolumn{1}{c|}{\centering -}  & \multicolumn{1}{c}{\centering $O(L(Td^2+T^2d))$} & \multicolumn{1}{c}{\centering -} & \multicolumn{1}{c}{\centering -} & \multicolumn{1}{c}{\centering $O(|\mathcal{V}|d)$}\\
     \multirow{1}{*} {\centering {Retrieval-Augmented SeRec}} & \multicolumn{1}{c}{\centering $O(|\theta|)$} & \multicolumn{1}{c|}{\centering $O(|\phi|)$} & \multicolumn{1}{c}{\centering $O(L(Td^2+T^2d))$} & \multicolumn{1}{c}{\centering $O(kd+\frac{|\mathcal{V}|}{k}d)$} & \multicolumn{1}{c}{\centering $O(Kd)$} & \multicolumn{1}{c}{\centering $O(|\mathcal{V}|d)$}\\
\bottomrule
\end{tabular}

\label{table:complexity}
\end{table*}
\subsection{Inference and Complexity Analyses}
\label{sec:infer_complex}
In this section, we first introduce the model inference process and then analyze the model complexity in terms of both time and space in detail.
\subsubsection{Model Inference} During inference, RaSeRec first computes the user representation $\mathbf{h}$ via $\mathrm{SeqEnc}(\cdot)$ for the input user sequence $s_u$. Then, RaSeRec retrieves top-$K$ similar memories (\textit{i.e.,} $\{ {\mathbf{h}}_{k}\}_{k=1}^K$ and $\{ {\mathbf{v}}_{k}\}_{k=1}^K$) from the memory bank $\mathcal{M}$. 
After that, RaSeRec employs RAM that consists of dual-channel MHCA to weight and aggregate valuable patterns from both $\{ {\mathbf{h}}_{k}\}_{k=1}^K$ and $\{ {\mathbf{v}}_{k}\}_{k=1}^K$. 
We obtain the final retrieval-augmented user representation $\tilde{\mathbf{h}}$ by combining both the original user representation $\mathbf{h}$ and the augmented user ones of the dual channels \textit{i.e.,} $\mathbf{h}^{c_1}$ and $\mathbf{h}^{c_2}$. Lastly, we recommend the top-$N$ items based on the dot product scores between $\tilde{\mathbf{h}}$ and item embedding matrix $\mathbf{V}$.
\subsubsection{Complexity Analyses} Table \ref{table:complexity} summarizes the space and time complexity of three SeRec paradigms, where we compare the complexity of RaSeRec with the Vanilla and SSL-Augmented SeRec to manifest the efficiency of RaSeRec. For space complexity, RaSeRec needs only $O(|\phi|+|\theta|)$, where $\phi$ and $\theta$ represent the trainable parameters of the RAM and sequence encoder, respectively. Generally, $|\phi|$ is far below than $|\theta|$. 
As such, the analytical space complexity of RaSeRec is in the same magnitude as the Vanilla and SSL-Augmented one ($O(|\theta|)$). In practice, taking the Amazon Sports dataset as an example, the space occupation of RaSeRec is merely 0.05 times larger than theirs.
As recommender systems mainly focus on the time cost of online serving, we analyze the time complexity during inference. The time complexity for encoding user representations is $O(L(Td^2+T^2d))$, where $T$ is the maximum sequence length and $d$ is the hidden size.
In addition to encoding, RaSeRec also needs to retrieve memories and augment user representation. For retrieval, RaSeRec instantiates Faiss using Inverted File Indexing (IVF), whose retrieval complexity is $O(kd+\frac{|\mathcal{V}|}{k}d)$, where $k$ denotes the number of clusters and we set the number of clusters accessed as 1.
On the other hand, the time complexity of augmenting user representation is $O(Kd)$. Finally, these three SeRec paradigms need to compute dot product scores over all items to make recommendations, resulting in a time complexity $O(|\mathcal{V}|d)$. 
The time complexity of additional retrieval and augmentation processes (\textit{i.e.,} $O(kd+\frac{|\mathcal{V}|}{k}d + Kd)$) is much lower than the encoding and recommendation processes (\textit{i.e.,} $O(L(Td^2+T^2d)+|\mathcal{V}|d)$). 
Specifically, $k+\frac{|\mathcal{V}|}{k}+K$ is on the order of thousands, while $|\mathcal{V}|$ is on the order of tens of thousands.  
Therefore, the analytical time complexity of RaSeRec is of the same magnitude as the Vanilla and SSL-Augmented SeRec.

\section{Experiments}
\label{sec:experiment}
To verify the effectiveness of RaSeRec, we conduct extensive experiments on three benchmark datasets to answer the following research questions (\textbf{RQs}): 
\begin{itemize}[leftmargin=*]
    \item \textbf{RQ1:} How does RaSeRec perform compared with \textit{state-of-the-art} SeRec models? 
    \item \textbf{RQ2:} How much gain can RaSeRec bring to the existing base backbones? 
    \item \textbf{RQ3:} Can RaSeRec improve long-tailed recommendation?
    \item \textbf{RQ4:} Can RaSeRec alleviate preference drift?
    \item \textbf{RQ5:} How do different partitions of the memory bank contribute to RaSeRec's performance?
    \item \textbf{RQ6:} Can RaSeRec perform robustly against the data noise issue?
    \item \textbf{RQ7:} How does the performance of RaSeRec vary with different hyper-parameter values? (Appendix \ref{app:para_sensi})
    \item \textbf{RQ8:} Whether RaSeRec benefits both high-frequency and low-frequency users? (Appendix \ref{app:user_freq})
\end{itemize}

\begin{table*}[t]
\tabcolsep=0.3cm
\renewcommand\arraystretch{1.0}
\footnotesize
\centering
\caption{Performance comparison (@5), where the best results are boldfaced and the second-best ones are underlined.}
\begin{tabular}{c|cc|cc|cc|cc}
\toprule
\textbf{Dataset} & \multicolumn{2}{c|}{Beauty} & \multicolumn{2}{c|}{Sports} & \multicolumn{2}{c|}{Clothing} & \multicolumn{2}{c}{Average} \\ \hline
Method  & HR@5    & NDCG@5     & HR@5     & NDCG@5       & HR@5       & NDCG@5   & HR@5       & NDCG@5      \\ \hline\hline
PopRec             &     0.0072         &     0.0040        &     0.0055         &    0.0040           &   0.0030        &   0.0018   &  0.0052         &      0.0033    \\
BPR-MF       &      0.0120        &      0.0065       &    0.0092          &     0.0053          &   0.0067        &  0.0052     &     0.0093      &      0.0057    \\
\hline
GRU4Rec         &    0.0164          &      0.0086       &        0.0137      &     0.0096          &    0.0095       &    0.0061      &  0.0132         &         0.0081       \\
Caser        &    0.0259          &   0.0127          &     0.0139         &    0.0085           &    0.0108       &   0.0067           &      0.0169     &       0.0093     \\
SASRec      &   0.0365           &   0.0236          &     0.0218         &      0.0127         &   0.0168        &   0.0091          &   0.0250        &    0.0151         \\
BERT4Rec     &      0.0193        &      0.0187       &      0.0176        &     0.0105          &    0.0125       &    0.0075        &    0.0165       &        0.0122      \\
$\rm S^3Rec_{MIP}$        &     0.0327         &      0.0175       &      0.0157        &     0.0098          &      0.0163     &   0.0101     &      0.0216     & 0.0125                 \\
\hline
CL4SRec     &       0.0394       &       0.0246      &      0.0238        &     0.0146          &    0.0163       &      0.0098             &    0.0265       &   0.0163    \\
CoSeRec      &     0.0463         &   0.0303          &   0.0278          &       0.0188        &     0.0161      &        0.0108            &     0.0301      & 0.0200     \\
ICLRec       &      0.0469        &     0.0305        &       0.0271       &     0.0182          &     0.0165     &      0.0104          &  0.0302         &     0.0197     \\
DuoRec      &        0.0541      &        0.0337     &        \underline{0.0315}      &      \underline{0.0196}         &     0.0191      &  0.0109        &    \underline{0.0349}       &      0.0214    \\
MCLRec       &      \underline{0.0552}        &     \underline{0.0347}        &      0.0294        &      0.0187         &      \textbf{0.0197}     &    \underline{0.0110}        &      0.0348     &      \underline{0.0215}        \\
\hline
RaSeRec (Ours)          &   \textbf{0.0569}   &    \textbf{0.0369}     &       \textbf{0.0331}       &     \textbf{0.0211}          &    \underline{0.0194}       &    \textbf{0.0118}  &     \textbf{0.0365}      &    \textbf{0.0233}  \\ \hline \hline
\%Improv.      &       3.08\%       &     6.34\%        &    5.08\%       &       7.65\%        &     -0.02\%      &  7.27\%   &     4.58\%      &   8.37\%  \\ \hline
$p$-value.       &   8$e^{-4}$           &   3$e^{-5}$           &      1$e^{-4}$        &        1$e^{-5}$       &     -      &   2$e^{-3}$   &   4$e^{-4}$         &   5$e^{-6}$   \\ 
\bottomrule
\end{tabular}

\label{table:overall_5}
\vspace{-0.15cm}
\end{table*}
\begin{table*}
\centering
\footnotesize
\tabcolsep=0.23cm
\renewcommand\arraystretch{1.05}
\caption{Performance comparison in terms of different sequential recommendation model backbones.}
\begin{tabular}{c|l|cccc|cccc}
\toprule

 \multirow{2}{*}{ \centering Backbone} & \multirow{2}{*}{\centering Model} & \multicolumn{4}{c|}{Beauty}&\multicolumn{4}{c}{Sports}\\ \cline{3-10}
  &   & HR@5 & HR@10 & NDCG@5 & NDCG@10 & HR@5 & HR@10 & NDCG@5 & NDCG@10  \\ \hline
  \rowcolor{gray!12}
  \cellcolor{white!20} \multirow{4}{*}{\centering GRU4Rec}& Base & 0.0164 & 0.0365  &  0.0086 & 0.0142 &  0.0137 & 0.0274 & 0.0096 &  0.0137   \\
  &  CL4SRec & 0.0314 & 0.0506 &  0.0196 & 0.0257 & 0.0196  & 0.0317 & 0.0126 & 0.0165\\ 
  & DuoRec & \underline{0.0443} & \underline{0.0691} & \underline{0.0296}  & \underline{0.0377} & \underline{0.0248}  & \underline{0.0386} & \underline{0.0162} &  \underline{0.0206}  \\
  & RaSeRec & \textbf{0.0469} & \textbf{0.0718} & \textbf{0.0322}  & \textbf{0.0403} & \textbf{0.0261}  & \textbf{0.0400} &  \textbf{0.0173} &   \textbf{0.0218}  \\
\hline
  \rowcolor{gray!12}
  \cellcolor{white!20} \multirow{4}{*}{\centering BERT4Rec} & Base & 0.0193 & 0.0401 &  0.0187 & 0.0254 & 0.0176  &  0.0326 & 0.0105 & 0.0153\\ 
  &  CL4SRec & 0.0322 & 0.0499 &  0.0195 & 0.0253 & 0.0226  & 0.0352 & 0.0132 & 0.0173\\ 
  & DuoRec & \underline{0.0505} & \underline{0.0759} &  \underline{0.0318} & \underline{0.0400} &  \underline{0.0275} & \underline{0.0436} & \underline{0.0170} &   \underline{0.0222}  \\
  & RaSeRec & \textbf{0.0534} & \textbf{0.0787} &  \textbf{0.0344} & \textbf{0.0426} & \textbf{0.0288}  & \textbf{0.0451} & \textbf{0.0182} &  \textbf{0.0236}   \\
\bottomrule
\end{tabular}

\label{table:plug_and_play}
\vspace{-0.15cm}
\end{table*}
\subsection{Experimental Settings}
\subsubsection{Datasets} We conduct extensive experiments on three benchmark datasets collected from Amazon \cite{DBLP:conf/sigir/McAuleyTSH15}. 
We adopt three sub-categories, \textit{i.e.,} Beauty, Sports, and Clothing\footnote{http://jmcauley.ucsd.edu/data/amazon/}, with different sparsity degrees. 
Following \cite{DBLP:conf/icdm/KangM18,DBLP:conf/cikm/SunLWPLOJ19}, we discard users and items with fewer than 5 interactions. The statistics of datasets after preprocessing are provided in Appendix \ref{app:datasets}.
\subsubsection{Baselines} To verify the effectiveness of RaSeRec, we compare it with the following three groups of competitive models: \textbf{(1) Non-sequential recommendation models (Non-SeRec)} include PopRec and BPR-MF \cite{DBLP:conf/uai/RendleFGS09}; 
\textbf{(2) Vanilla sequential recommendation models (Vanilla SeRec)} include GRU4Rec \cite{DBLP:journals/corr/HidasiKBT15}, Caser \cite{DBLP:conf/wsdm/TangW18}, SASRec \cite{DBLP:conf/icdm/KangM18}, BERT4Rec \cite{DBLP:conf/cikm/SunLWPLOJ19}, and $\rm S^3Rec_{MIP}$ \cite{DBLP:conf/cikm/ZhouWZZWZWW20}; 
and { \bf (3) SSL-Augmented sequential recommendation models (SSL-Augmented SeRec)} include CL4SRec \cite{DBLP:conf/icde/XieSLWGZDC22}, CoSeRec \cite{DBLP:journals/corr/abs-2108-06479}, ICLRec \cite{DBLP:conf/www/ChenLLMX22}, DuoRec \cite{DBLP:conf/wsdm/QiuHYW22}, and MCLRec \cite{DBLP:conf/sigir/QinYZFZ0LS23}. Detailed descriptions of each baseline model can be seen in Appendix \ref{app:baseline}.
\subsubsection{Implementation and Evaluation} We use Recbole \cite{DBLP:conf/cikm/ZhaoHPYZLZBTSCX22} to implement baselines. 
For models with learnable item embedding, we set the hidden size $d=64$. For each baseline, we follow the suggested settings reported in their original paper to set hyper-parameters. 
More implementation details can be seen in Appendix \ref{app:imple}.
Following the leave-one-out strategy, we hold out the last interacted item of each user sequence for testing, the second last item for validation, and all the earlier ones for training.
We measure the recommendation performance via HR and NDCG (see Appendix \ref{app:metrics}), where the cut-off position (@$N$) is set as 5 and 10. 
\subsection{Comparison with Baselines (RQ1)}
\label{sec:compari_base}
Table \ref{table:overall_5} and \ref{table:overall_10} summarize the performance of all baselines and RaSeRec \textit{w.r.t.} @5 and @10, respectively. From the experimental results, we observe that: 
\textbf{(1)} The performance of the Non-SeRec models is unsurprisingly worse than the vanilla SeRec models, indicating the necessity of modeling sequential patterns for capturing more accurate user preferences. On the other hand, SSL-Augmented SeRec models perform better than those without SSL, indicating the superiority of supplementing the recommendation task with self-supervised learning.
\textbf{(2)} Benefiting from the retrieval-augmented mechanism, RaSeRec considerably outperforms all baseline models in 16 out of 18 cases with $p$-value less than 0.005, demonstrating the improvement of RaSeRec over the SOTA baselines is statistically significant. The reasons are mainly twofold: 
\textbf{(i)} By sustainably maintaining a dynamic memory bank, RaSeRec can quickly adapt to preference drifts. 
\textbf{(ii)} By retrieving memories, RaSeRec can explicitly recall long-tailed patterns that may be overwhelmed by head ones. 
These above two innovative designs contribute to considerable improvements together.
\subsection{Improving Base Backbones (RQ2)}
\label{sec:impro_base}
RaSeRec is a model-agnostic retrieval-augmented SeRec paradigm that can enhance various base models. 
To verify this, instead of using the default Self-attentive backbone SASRec \cite{DBLP:conf/icdm/KangM18}, we evaluate RaSeRec using other backbones, \textit{i.e.,} RNN-based GRU4Rec \cite{DBLP:journals/corr/HidasiKBT15} and Bert-based BERT4Rec \cite{DBLP:conf/cikm/SunLWPLOJ19}. 
As the average performance of DuoRec and MCLRec is close, we mainly compare with DuoRec in the following.
As shown in Table \ref{table:plug_and_play}, RaSeRec performs best with different backbones, indicating that RaSeRec has a good generalization ability. 
The impressive performance gain demonstrates the effectiveness of our retrieval-augmented mechanism, even though we just add a very small set of parameters (about 5\%).  
In summary, RaSeRec possesses the virtue of the plug-and-play property, which can enhance various given base models with retrieval augmentation while leaving the base model backbone unchanged.
\begin{figure}[t]
    \centering  
     \subfigure[Results on the Beauty dataset \textit{w.r.t.} NDCG@5.]{
        \includegraphics[width=1.00\linewidth]{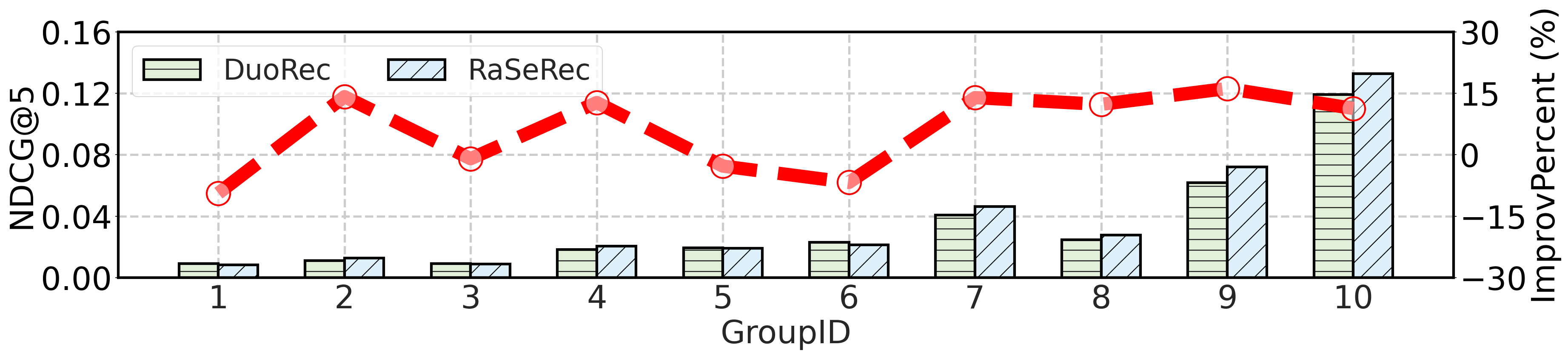}\label{fig:beauty_item_fren}}
    \subfigure[Results on the Sports dataset \textit{w.r.t.} NDCG@5.]{
        \includegraphics[width=1.00\linewidth]{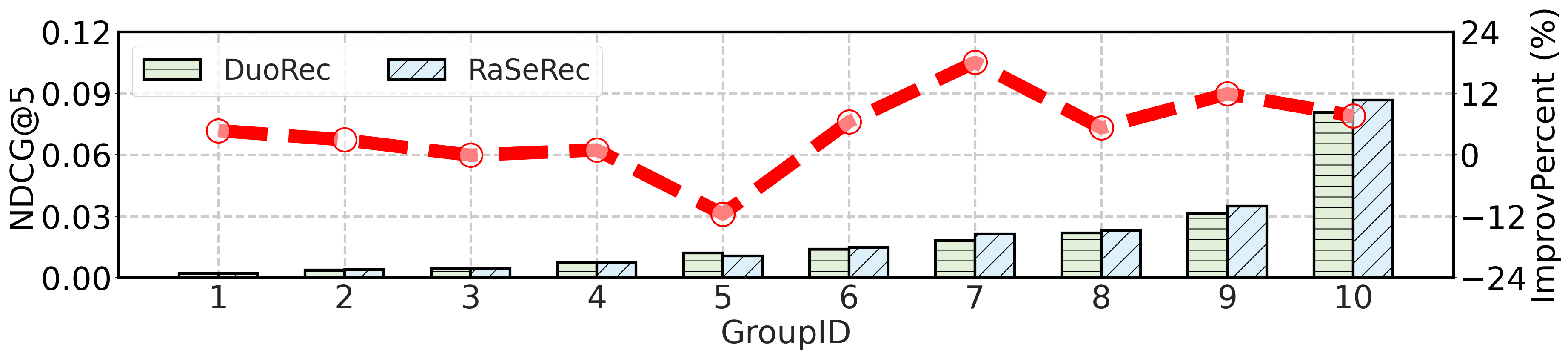}\label{fig:sports_item_fren}}
    \caption{Performance comparison over different item groups between RaSeRec and DuoRec. The bar represents NDCG@5, while the line represents the performance improvement percentage of RaSeRec compared with DuoRec.}
    \label{fig:item_fren}
    \vspace{-0.25cm}
\end{figure}
\begin{figure*}[t]
    \centering  
     \subfigure[Results on the Beauty dataset \textit{w.r.t.} HR@5 and NDCG@5.]{
        \includegraphics[width=1.00\linewidth]{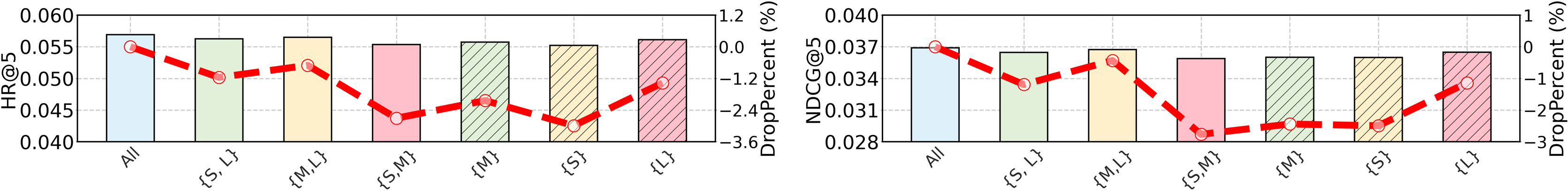}\label{fig:beauty_abla_memo}}
    \subfigure[Results on the Sports dataset \textit{w.r.t.} HR@5 and NDCG@5.]{
        \includegraphics[width=1.00\linewidth]{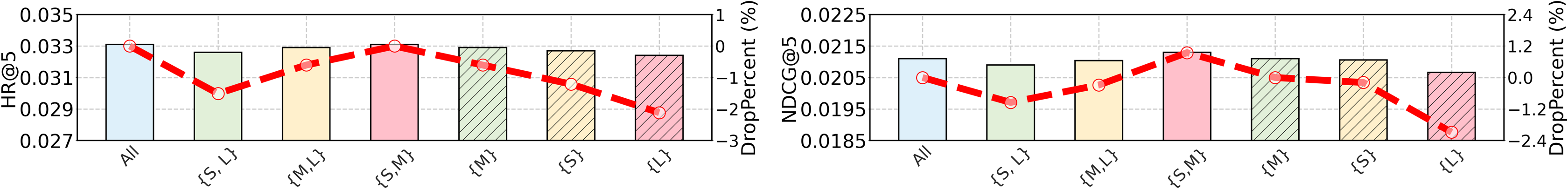}\label{fig:sports_abla_memo}}
    \subfigure[Results on the Clothing dataset \textit{w.r.t.} HR@5 and NDCG@5.]{
        \includegraphics[width=1.00\linewidth]{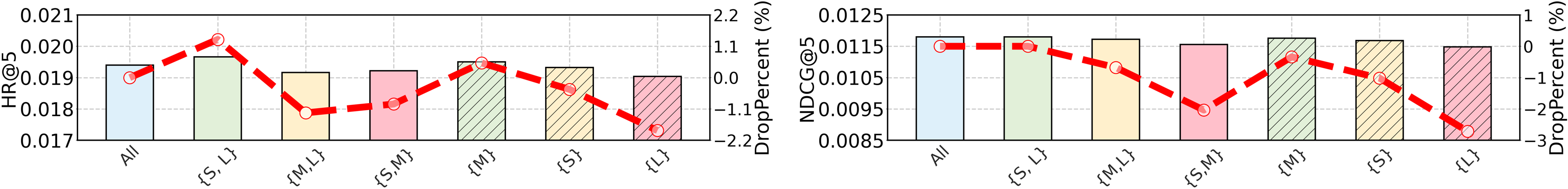}\label{fig:clothing_abla_memo}}
    \vspace{-0.15cm}
    \caption{Performance comparison when ablating different partitions of the memory bank $\mathcal{M}$. The bar represents HR@5 or NDCG@5, while the line represents the percentage of performance degradation compared to the {``All''}.}
    \label{fig:abla_memo}
    \vspace{-0.15cm}
\end{figure*}
\subsection{Long-tailed Recommendation (RQ3)}
\label{sec:long_tail}
As stated in Section \ref{sec:intro}, existing SeRec models make it hard to recall long-tailed patterns via implicit memories due to skewed supervised signal distribution. 
To verify whether RaSeRec is of promise in solving this issue, we split the test user sequences into 10 groups based on the target item’s popularity measured by the number of interactions, meanwhile keeping the total number of test user sequences in each group the same. 
And the larger the GroupId, the more popular the target item. The results are shown in Figure \ref{fig:item_fren}.
From the results, we observe that: \textbf{(1)} Compared with DuoRec, the performance improvements of RaSeRec mostly come from accurately recommending long-tailed and head items. This verifies that retrieval-augmented user representation well recalls both long-tailed and head patterns.
\textbf{(2)} We observe that RaSeRec performs relatively worse in recommending general items, \textit{i.e.,} $5$-th and $6$-th groups. We think the main reason is that the relevant memories are not retrieved while the noisy memories are introduced. 
How to effectively retrieve relevant knowledge is also an open challenge in RAG \cite{DBLP:conf/emnlp/ZhaoLZHCHZ24,DBLP:journals/corr/abs-2311-08377,DBLP:conf/acl/LiLCX024}. In summary, we suggest performing retrieval augmentation when recommending the long-tailed and head items.
\begin{table}[t]
  \renewcommand\arraystretch{1.05}
  \tabcolsep=0.225cm
  \caption{Model performance \textit{w.r.t.} preference drifts.}
  \centering
  \footnotesize
  \begin{tabular}{c|l|cc|cc}
    \toprule
    \multicolumn{2}{c|}{ \multirow{1}{*}[+0.2ex]{\centering \bf Metrics}} & 
    \multicolumn{2}{c|}{ \multirow{1}{*}[+0.2ex]{\centering \bf HR}} & 
    \multicolumn{2}{c}{ \multirow{1}{*}[+0.2ex]{\centering \bf NDCG}} \\
    \cline{1-6}
     \multicolumn{1}{c|}{ \multirow{1}{*}[-0.2ex]{\centering \bf Datasets}} & \multicolumn{1}{c|}{ \multirow{1}{*}[-0.2ex]{\centering \bf Drifts}} &\multirow{1}{*}[-0.2ex]{\centering \bf @5} & \multirow{1}{*}[-0.2ex]{\centering \bf @10} & \multirow{1}{*}[-0.2ex]{\centering \bf @5} & \multirow{1}{*}[-0.2ex]{\centering \bf @10}  \\
     
    \cline{1-6}
     \multicolumn{1}{c|}{ \multirow{4}{*} {\centering \bf Beauty}} & \textbf{(A) Full} & \textbf{0.0569} & \textbf{0.0860} & \textbf{0.0369} & \textbf{0.0463}  \\
      & (B) 10\% & \underline{0.0560} & {0.0842} & \underline{0.0360} & \underline{0.0450} \\
      & (C) 20\% & {0.0550} & \underline{0.0843} & 0.0355 & {0.0449} \\
      & (D) 30\% & 0.0542 & \underline{0.0843} & 0.0350 & 0.0447\\

      \cline{1-6}
     \multicolumn{1}{c|}{ \multirow{4}{*} {\centering \bf Sports}} & \textbf{(A) Full}& \textbf{0.0331} & \textbf{0.0497} & \textbf{0.0211} & \textbf{0.0264} \\
      & (B) 10\% & 0.0326 & \underline{0.0492} & \underline{0.0208} & \underline{0.0261}\\
      & (C) 20\% & 0.0322 & 0.0490 & {0.0204} & {0.0258} \\
      & (D) 30\% & \underline{0.0328} & {0.0490} & \underline{0.0208} & {0.0260} \\
    \bottomrule
  \end{tabular}
  \label{table:pd}
  \vspace{-0.2cm}
\end{table}

\subsection{Alleviating Preference Drift (RQ4)}
As stated in Section \ref{sec:intro}, existing SeRec models trained on past data may recommend undesirable items due to preference drift. 
To verify whether RaSeRec is promising in alleviating this issue, we simulate this scenario by removing a certain ratio (\textit{i.e.,} 10\%, 20\%, and 30\%) of the preference data with the latest timestamp from the memory bank. 
And, ``Full'' denotes using all preference data in the memory bank to remedy preference drift.
From the results shown in Table \ref{table:pd}, we find that (1) In general, the larger the preference drifts, the worse the model performance; 
(2) ``Full'' performs best as it can remedy preference drift by maintaining a dynamic memory bank with the latest preference data; 
and (3) In a few cases, large preference drift is not necessarily worse performance, \textit{e.g.,} (D) on the sports dataset. The main reason is different datasets have different preferences towards short-, medium-, and long-term memories, which also greatly affects performance (more details can be seen in \S\ref{app:abla_memo}). 
In a nutshell, the above observations verify that RaSeRec can alleviate preference drift effectively.
\begin{table}[t]
  \centering
  \footnotesize
  \renewcommand\arraystretch{1.0}
  \tabcolsep=0.125cm
  \caption{\centering{Statistics of each partition.}}
  \begin{tabular}{c|c|c|c}
    \toprule
    Dataset & \#len.range & \#avg.length & \%proportion \\
    \hline
    \multirow{3}{*}{\centering Beauty} & <3 & 1.50 & 34.03\% \\
     &  3-6 & 4.15 & 31.18\% \\
     & >6 & 18.88 & 34.79\% \\ \hline
     \multirow{3}{*}{\centering Sports} & <3 & 1.50 & 37.56\% \\
     &  3-6 & 4.14 & 33.98\% \\
     & >6 & 15.31 & 28.46\% \\ \hline
     \multirow{3}{*}{\centering Clothing} & <2 & 1.00 & 24.54\% \\
     &  2-3 & 2.38 & 39.38\%\\
     & >3 & 8.06 & 36.08\%\\ 
    \bottomrule
  \end{tabular}
  \label{table:dataset_part}
  \vspace{-0.15cm}
\end{table}
\subsection{Study on Memory Bank (RQ5)}
\label{app:abla_memo}
To investigate the influence of different partitions of the memory bank on the model performance, we divide the memory bank $\mathcal{M}$ into three partitions according to the length of user sequences. The statistics of each partition are summarized in Table \ref{table:dataset_part}. 
For convenience, we use $\mathrm{S}$, $\mathrm{M}$, and $\mathrm{L}$ to represent the \underline{S}hort-term, \underline{M}edium-term, and \underline{L}ong-term memory bank, respectively. For example, on the beauty dataset, we treat user sequences with length less than 3, those between 3 and 6, and those greater than 6 as $\mathrm{S}$, $\mathrm{M}$, and $\mathrm{L}$ memory bank respectively, where the proportion of different partitions accounts for about 1/3.
The experimental results are shown in Figure \ref{fig:abla_memo}. ``All'' denotes using all parts of the memory bank, including $\mathrm{S}$, $\mathrm{M}$, and $\mathrm{L}$, while $\{\mathrm{S}, \mathrm{L}\}$ denotes using parts of the memory bank, \textit{i.e.,} $\mathrm{S}$ and $\mathrm{L}$, and so on.
\begin{figure}[t]
    \centering  
     \subfigure[Results on Beauty \textit{w.r.t.} HR@5 and NDCG@5.]{
        \includegraphics[width=0.99\linewidth]{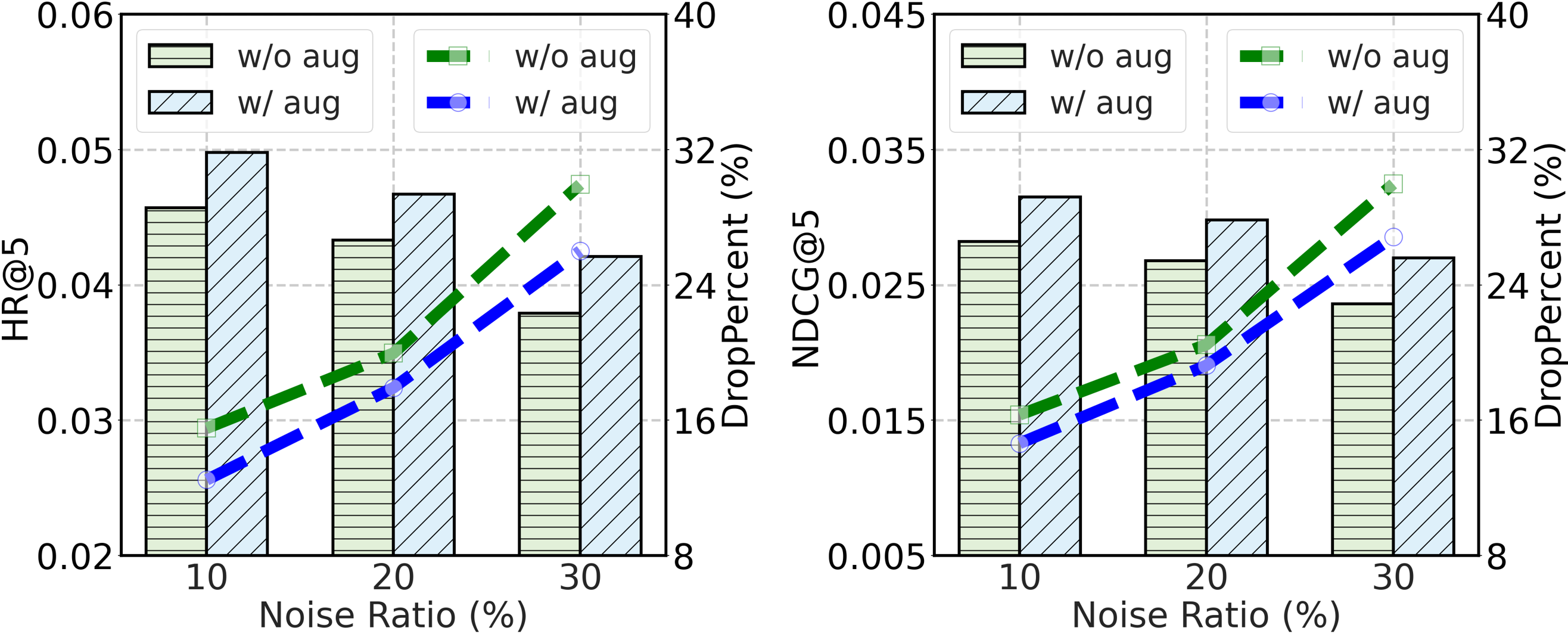}\label{fig:beauty_robust_noise}}
    \subfigure[Results on Sports \textit{w.r.t.} HR@5 and NDCG@5.]{
        \includegraphics[width=0.99\linewidth]{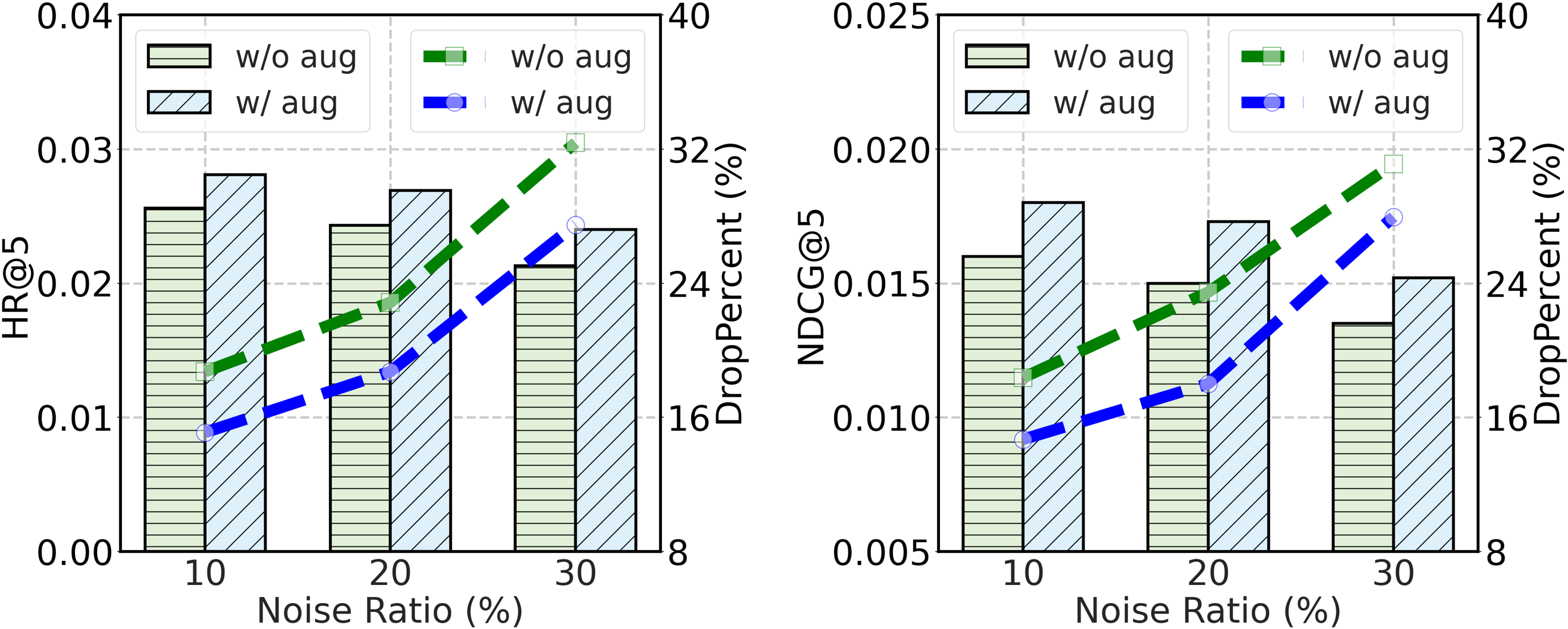}\label{fig:sports_robust_noise}}
    \caption{Model performance \textit{w.r.t.} noise ratio on Beauty and Sports datasets. The bar represents HR@5 or NDCG@5, while the line represents the percentage of performance degradation compared to the test user sequences without added noise.}
    \vspace{-0.15cm}
    \label{fig:robust_noise}
\end{figure}
From the experimental results, we observe that different datasets show different preferences toward the combinations of the partitions. Specifically, the beauty dataset prefers ``All''; the sports dataset prefers $\{\mathrm{S}, \mathrm{M}\}$; the clothing dataset prefers $\{\mathrm{S}, \mathrm{L}\}$. We think the main reason is associated with \#avg.length. 

As for the beauty dataset, its average length is longer than the other two datasets and it prefers the $\mathrm{L}$ partition. Specifically, the models with $\mathrm{L}$ partition largely outperform those without $\mathrm{L}$ partition, as shown in Figure \ref{fig:beauty_abla_memo}.
As for the sports and clothing datasets, their average length is shorter than that of the beauty dataset, and they perform relatively worse when only provided with $\mathrm{L}$ partition, as shown in Figure \ref{fig:sports_abla_memo} and \ref{fig:clothing_abla_memo}. Besides, we observe that the model equipped with parts of the memory bank (\textit{e.g.,} $\{\mathrm{S,M}\}$ is the best for the sports dataset and $\{\mathrm{S,L}\}$ is the best for the clothing dataset) even outperforms that with all parts of the memory bank. 
The above observations demonstrate that not all memories are beneficial and some ones are harmful to recommendation performance. This is also a challenging issue \cite{DBLP:conf/sigir/CuconasuTSFCMTS24,DBLP:journals/corr/abs-2405-15556,DBLP:conf/acl/FangBNY0X24} in RAG that warrants further research. In a nutshell, we suggest trying different combinations or directly using all of the memories, because ``All'' usually leads to satisfactory performance. In this work, we use ``All'' by default.

\subsection{Robustness against Noise Data (RQ6)}
In real-world scenarios, recommender systems usually suffer from the noisy interaction issue \cite{DBLP:conf/sigir/Gao0HCZFZ22,DBLP:conf/sigir/TianXLYZ22}, \textit{e.g.,} a user clicked an item by mistake. 
As such, it is necessary to verify RaSeRec’s robustness against noisy data. To simulate this scenario, we train models with full training data and randomly add a certain ratio (\textit{i.e.,} 10\%, 20\%, and 30\%) of negative items into each test user sequence. 
The experimental results are shown in Figure \ref{fig:robust_noise}, where models with retrieval augmentation mention ``w/ aug'' otherwise ``w/o aug''.
From the experimental results, we observe that adding noises significantly degrades the performance of models with or without retrieval augmentation. However, ``w/ aug'' consistently outperforms ``w/o aug'' under any noise ratio in terms of HR@5 and NDCG@5. 
Moreover, the blue line is always below the green line, as shown in Figure \ref{fig:robust_noise}, demonstrating that ``w/ aug'' can endow the model backbone with more robustness against the noisy interactions during inference. 
For example, on the Beauty and Sports datasets, ``w/ aug'' with a 20\% noise ratio outperforms ``w/o aug'' with a 10\% noise ratio. Moreover, on the Sports dataset, ``w/ aug'' drops 14.7\% of its original performance, while ``w/o aug'' drops 18.4\%, in terms of NDCG@5, when both have a 10\% noise ratio. This fully implies that with retrieval augmentation, RaSeRec can retrieve useful sequential patterns as references to alleviate the influence of noisy interactions and make a suitable recommendation.
In a nutshell, we suggest performing retrieval augmentation to alleviate the negative influence of noisy interactions.

\section{Related Works}
\subsection{Retrieval-Augmented Generation}
Recently, RAG prevails in LLMs, introducing non-parametric knowledge into LLMs to supplement their incomplete, incorrect, or outdated parametric knowledge \cite{DBLP:journals/corr/abs-2312-10997}.
It has been shown highly effective in many research fields, not limited to natural language, including computer vision \cite{DBLP:conf/cvpr/XieSXZZZ23,DBLP:journals/corr/abs-2405-10311,rao2022does,rao2023dynamic}, speech \cite{DBLP:journals/corr/abs-2406-03714,DBLP:conf/icassp/WangSS00YS24}, code \cite{DBLP:conf/acl/LiLXY00023,DBLP:journals/corr/abs-2406-14497}, and graph \cite{DBLP:journals/corr/abs-2404-16130}. 
Although being well-studied in many fields, very limited works exploit RAG in recommender systems, let alone SeRec. A very recent one is RUEL \cite{DBLP:conf/cikm/WuGSPJ23}. 
However, it differs from our work in \textbf{(1)} it lacks retrieval training, and \textbf{(2)} it falls short in decoupling implicit and explicit memories, while our work proposes a model-agnostic collaborative-based pre-training and retrieval-augmented fine-tuning paradigm for SeRec that can be applied to any ID-based SeRec backbones and improve their performance. 
\subsection{Sequential Recommendation}
Pioneering attempts on SeRec are based on Markov Chain (MC) to model item-item transition relationships, \textit{e.g.,}  FPMC \cite{DBLP:conf/www/RendleFS10}. 
To capture long-term dependencies, some studies first adopt RNN for SeRec by modeling sequence-level item transitions, \textit{e.g.,} GRU4Rec \cite{DBLP:journals/corr/HidasiKBT15}. Simultaneously, CNN has been explored for SeRec by modeling the union-level sequential patterns \cite{DBLP:conf/wsdm/TangW18}. 
Recently, the success of self-attention models in NLP \cite{DBLP:conf/nips/VaswaniSPUJGKP17} has spawned a series of Transformer-based sequential recommendation models, \textit{e.g.,} SASRec \cite{DBLP:conf/icdm/KangM18} and BERT4Rec \cite{DBLP:conf/cikm/SunLWPLOJ19}.
Considering sparse supervised signals, SSL has been widely adopted in SeRec to alleviate this issue \cite{DBLP:conf/icde/XieSLWGZDC22,DBLP:conf/sigir/QinYZFZ0LS23,DBLP:conf/wsdm/QiuHYW22}.
Enlightened by generative models, some studies have attempted to apply LLMs to solve SeRec \cite{DBLP:conf/icde/ZhengHLCZCW24,DBLP:conf/kdd/LiWLFSSM23,DBLP:conf/ecom/LiZWXLM23,DBLP:conf/www/ZhengCQZ024}. 
Albeit studied for ages, almost all existing methods focus on exploiting implicit memories hidden in the model. 
This work is in an orthogonal direction. It exploits explicit memories in a retrieval-augmented manner and opens a new research direction for SeRec.
\section{Conclusion and Future Works}
In this work, we unveiled the limitations of the Vanilla SeRec and SSL-Augmented SeRec paradigms and explored the potential of RAG to solve them.
In particular, we propose a retrieval-augmented SeRec paradigm, which learns to refine user representation with explicit retrieved memories and adapt to preference drifts by maintaining a dynamic memory bank. 
Extensive experiments on three datasets manifest the advantage of RaSeRec in terms of recommendation performance, long-tailed recommendation, and alleviating preference drift.
Despite our innovations and improvements, we recognize some issues that warrant further study. 
In this work, we fixed the number of retrieved memories $K$ during training and inference. 
In Appendix \ref{app:para_sensi}, we have studied the impact of different $K$ values on the model performance. 
We find both too large and small $K$ values hurt the model performance. 
Furthermore, in Section \ref{sec:long_tail} and Appendix \ref{app:user_freq}, we have studied the impact of retrieval augmentation on different item groups and user groups. 
The experimental results show that retrieval augmentation may hurt the performance of some groups. 
These observations motivate us to develop an active retrieval augmentation mechanism for RaSeRec, \textit{i.e.,} retrieval when needed.

\bibliographystyle{ACM-Reference-Format}
\bibliography{sample-base}

\appendix

\section{Notations}
\label{app:notation}
Table \ref{table:notation} summarizes the main notations used in this paper, including some essential hyper-parameters.
\begin{table}[htbp]
\footnotesize
\renewcommand\arraystretch{1.1}
  \centering
    \caption{Notation table.}
    \begin{tabular}{cc}
    \toprule
    \textbf{Notation} & \textbf{Meaning} \\
    \hline
    $\mathcal{U}, \mathcal{V}$ & set of users and items\\
    $\theta, \phi$ & parameters of the backbone and RAM\\
    $\mathbf{V}$ & item embedding table\\
    $\mathbf{P}$ & position encoding matrix\\
    $\mathbf{h}, \mathbf{h}^{\prime}, \mathbf{h}^{\prime\prime}$ & user representation \\
    $\tilde{\mathbf{h}}, \mathbf{h}^{c_1}, \mathbf{h}^{c_2}$ & augmented user representation \\
    $\mathbf{v}_i$ & item embedding of $i$-th item\\
    $s_u$ & interaction sequence of user $u$\\
    $v_{t}^{(u)}$ & item interacted by user $u$ at timestamp $t$\\
    $L$ & number of Transformer layers \\
    $N$ & number of items recommended \\
    $T$ & maximum input sequence length\\
    $d$ & hidden size \\
    $K$ & number of memories to retrieve \\
    $k$ & number of clusters when building Faiss\\
    $\alpha, \beta$ & control coefficient \\
    \bottomrule
    \end{tabular}
  \label{table:notation}
\end{table}
\section{Datasets}
\label{app:datasets}
Table \ref{table:dataset} summarizes the detailed statistics of three benchmark datasets after preprocessing, where \#users, \#items, and \#inters represent the number of users, items, and interactions, \#avg.length denotes the average length of all user sequences, and `sparsity' calculates the proportion of entries in the user-item matrix without any interaction records.
\begin{table}[ht]
  \centering
  \footnotesize
  \renewcommand\arraystretch{1.1}
  \tabcolsep=0.125cm
  \caption{\centering{Statistics of the datasets.}}
  \begin{tabular}{lccccc}
    \toprule
    Dataset & \#users & \#items & \#inters & \#avg.length & sparsity\\
    \midrule 
    {Beauty} & 22363 & 12101 & 198502 &  8.88 & 99.93\%\\
    {Sports} & 35598 & 18357 & 296337 &  8.32 & 99.95\%\\
    {Clothing} & 39387 & 23033 & 278677 &  7.08 & 99.97\%\\
    \bottomrule
  \end{tabular}
  \label{table:dataset}
\end{table}
\section{Baselines}
\label{app:baseline}
We compare RaSeRec with the following three groups of competitive baseline models:

\noindent\textbf{(1) Non-sequential recommendation models (Non-SeRec):}
\begin{itemize}[leftmargin=*]
    \item PopRec is a non-personalized model based on the popularity of items to recommend.
    \item BPR-MF \cite{DBLP:conf/uai/RendleFGS09} uses matrix factorization to model users and items with BPR loss.
\end{itemize}
\noindent\textbf{(2) Vanilla sequential recommendation models (Vanilla SeRec):}
\begin{itemize}[leftmargin=*]
    \item GRU4Rec \cite{DBLP:journals/corr/HidasiKBT15} uses the Gate Recurrent Unit (GRU) to model user sequence.
    \item Caser \cite{DBLP:conf/wsdm/TangW18} applies Convolutional Neural Networks (CNN) to model union-level sequential patterns for SeRec.
    \item SASRec \cite{DBLP:conf/icdm/KangM18} models user sequences via self-attention mechanism.
    \item BERT4Rec \cite{DBLP:conf/cikm/SunLWPLOJ19} applies the bi-directional self-attention mechanism with masked item training to model user sequence.
    \item $\rm S^3Rec_{MIP}$ \cite{DBLP:conf/cikm/ZhouWZZWZWW20} learns the correlation among items, attributes and etc. Here, we adopt the Mask Item Prediction (MIP) variant.
\end{itemize}
\noindent\textbf{(3) SSL-Augmented sequential recommendation models (SSL-Augmented SeRec):}
\begin{itemize}[leftmargin=*]
    \item CL4SRec \cite{DBLP:conf/icde/XieSLWGZDC22} proposes item crop, mask, and reorder to construct different views of user sequences.
    \item CoSeRec \cite{DBLP:journals/corr/abs-2108-06479} propose two robust data augmentations (\textit{i.e.,} item substitute and insert), to create high-quality views for SSL.
    \item ICLRec \cite{DBLP:conf/www/ChenLLMX22} first mines the latent users' intents and creates SSL signals between sequences and intents.
    \item DuoRec \cite{DBLP:conf/wsdm/QiuHYW22} proposes model-level augmentation with different sets of dropout masks on the model backbone. 
    \item MCLRec \cite{DBLP:conf/sigir/QinYZFZ0LS23} designs learnable augmenters to generate views and applies meta-learning to guide the training of augmenters. 
\end{itemize} 
%

%
\begin{table*}[t]
\tabcolsep=0.3cm
\renewcommand\arraystretch{1.}
\footnotesize
\centering
\caption{Performance comparison (@10), where the best results are boldfaced and the second-best ones are underlined.}
\begin{tabular}{c|cc|cc|cc|cc}
\toprule
\textbf{Dataset} & \multicolumn{2}{c|}{Beauty} & \multicolumn{2}{c|}{Sports} & \multicolumn{2}{c|}{Clothing} & \multicolumn{2}{c}{Average} \\ \hline
Method  & HR@10    & NDCG@10     & HR@10     & NDCG@10       & HR@10       & NDCG@10   & HR@10       & NDCG@10      \\ \hline\hline
PopRec             &   0.0114           &     0.0053        &      0.0090        &        0.0051       &     0.0052      &  0.0025    &      0.0085     &     0.0043     \\
BPR-MF       &     0.0299         &      0.0122       &       0.0188       &       0.0083        &    0.0094       &    0.0069    &      0.0194     &     0.0091     \\
\hline
GRU4Rec         &    0.0365          &    0.0142         &       0.0274       &        0.0137       &      0.0165     &     0.0083     &      0.0268     &    0.0121            \\
Caser        &       0.0418       &       0.0253      &     0.0231         &     0.0126          &      0.0174     &         0.0098     &     0.0274      &    0.0159        \\
SASRec      &        0.0627      &      0.0281       &    0.0336          &     0.0169          &     0.0272      &   0.0124          &     0.0412      &   0.0191          \\
BERT4Rec     &      0.0401        &       0.0254      &        0.0326      &        0.0153       &       0.0208    &     0.0102       &     0.0312      &    0.0170          \\
$\rm S^3Rec_{MIP}$        &      0.0591        &   0.0268          &        0.0265      &       0.0135        &      0.0237     &   0.0132     &       0.0364    &     0.0178             \\
\hline

CL4SRec     &      0.0667        &         0.0334    &      0.0392        &    0.0195           &      0.0274     &       0.0134            &      0.0444     &   0.0221    \\
CoSeRec      &     0.0705         &     0.0381        &    0.0422          &   0.0234            &   0.0248        &       0.0136             &     0.0458      &  0.0250    \\
ICLRec       &         0.0729     &       0.0389      &       0.0433       &        0.0234       &       0.0256    &        0.0134        &      0.0473     &    0.0252      \\
DuoRec      &        0.0834     &        0.0431     &         \underline{0.0479}     &         \underline{0.0248}      &      0.0293     &  0.0142              &   0.0535        &      \underline{0.0274}    \\
MCLRec       &       \underline{0.0844}       &       \underline{0.0437}      &    0.0462          &        0.0241       &      \textbf{0.0306}     &    \underline{0.0144}        &    \underline{0.0537}       &      \underline{0.0274}        \\
\hline
RaSeRec (Ours)          &       \textbf{0.0860}       &     \textbf{0.0463}        &        \textbf{0.0497}      &   \textbf{0.0264}            &    \underline{0.0301}       &   \textbf{0.0152}   &      \textbf{0.0553}     &   \textbf{0.0293}   \\ \hline \hline
\%Improv.      &       1.90\%       &      5.95\%       &     3.76\%         &       6.45\%        &      -0.02\%     &  5.56\%  &     2.98\%      &  6.93\%   \\ \hline
$p$-value.       &      9$e^{-3}$        &       5$e^{-5}$      &     4$e^{-4}$         &       1$e^{-4}$        &     -      &  5$e^{-4}$   &      2$e^{-4}$     &   1$e^{-6}$   \\ 
\bottomrule
\end{tabular}

\label{table:overall_10}
\end{table*}
\section{Metrics}
\label{app:metrics}
We rank the prediction results over the whole item set $\mathcal{V}$ and employ Hit Rate (HR) and Normalized Discounted Cumulative Gain (NDCG) to measure recommendation performance. Assuming the cut-off position is $N$, HR@$N$ and NDCG@$N$ can be defined as:
\begin{itemize}[leftmargin=*]
    \item \textbf{HR@\textit{N}} measures whether the target item of the test user sequence appears in the first $N$ recommended items:
    \begin{equation}
        \text{HR@}N = \frac{1}{|\mathcal{U}|} \sum_{u\in \mathcal{U}} \delta(v_{t+1}\in R(u)),
    \end{equation}
    where $\delta(\cdot)$ is an indicator function; $v_{t+1}$ denotes the target item; $R(u)$ denotes the top-$N$ recommended list of user $u$.
    \item \textbf{NDCG@\textit{N}} further measures the ranking quality. It logarithmically discounts the position:  
    \begin{equation}
        \begin{split}
            &\text{NDCG@}N = \frac{1}{|\mathcal{U}|} \sum_{u\in \mathcal{U}} \sum_{j=1}^{N} \\
            &\delta(v_{t+1}=R(u)_j) \frac{1}{\log_2(j+1)},
        \end{split}
    \end{equation}
    where $R(u)_j$ denotes the $j$-th recommended item for user $u$.
\end{itemize}
\begin{table}[t]
\footnotesize
\renewcommand\arraystretch{1.1}
  \centering
  \caption{Hyperparameter settings, where * denotes different hyperparameter settings for different datasets.}
    \begin{tabular}{cc}
    \toprule
    \textbf{Configuration} & \textbf{Value} \\
    \hline
    number of Transformer layers $L$ & 2\\
    number of heads & 2\\
    hidden size $d$ & 64\\
    maximum sequence length $L$ & 50\\ 
    number of clusters $k$ & 128 \\
    number of clusters accessed & 1 \\
    control coefficient $\alpha$\textsuperscript{*} & $[0.0, 1.0]$\\
    control coefficient $\beta$\textsuperscript{*} & $[0.0, 1.0]$\\
    number of retrieved memories $K$\textsuperscript{*} & $[5, 55]$\\
    learning rate & 0.001\\
    optimizer          & Adam \\
    mini-batch size          & 1024 \\
    \bottomrule
    \end{tabular}
    
  \label{table:hyper}
\end{table}

\section{Implementation Details} 
\label{app:imple}
We use the Adam optimizer \cite{DBLP:journals/corr/KingmaB14} to optimize model parameters with the learning rate of 0.001, the mini-batch size of 1024, $\beta_1=0.9$, and $\beta_2=0.999$, where the maximum number of epochs is set to 100 for both pre-training and fine-tuning stages.
We train models with the early stopping strategy that interrupts training if the HR@10 result on the validation set continues to drop until 10 epochs. 
For RAM, we tune $\alpha$, $\beta$ and $K$ within the ranges of $\{0.0, 0.1, 0.2, ..., 1.0\}$, $\{0.0, 0.1, 0.2, ..., 1.0\}$ and $\{5, 10, 15, ..., 55\}$, respectively, referring to Appendix \ref{app:para_sensi} for more details. 
As for Faiss, we set the number of clusters $k$ to 128, set the number of clusters accessed as 1, and adopt the cosine similarity as the measurement to retrieve relevant memories.
As for the sequence encoder, we adopt a two-layer Transformer and set the head number as 2 for each self-attention block, where the hidden size $d$ and the maximum sequence length $L$ are set to 64 and 50 respectively.
Additionally, for Non-SeRec and Vanilla SeRec methods, we use results reported by DuoRec \cite{DBLP:conf/wsdm/QiuHYW22} as these methods already have generally acknowledged experimental results on the three benchmark datasets used in this paper. We ran each experiment five times and reported the average. Table \ref{table:hyper} summarizes the detailed hyper-parameter settings.
\begin{figure*}[t]
    \centering  
     \subfigure{
        \includegraphics[width=.309\linewidth]{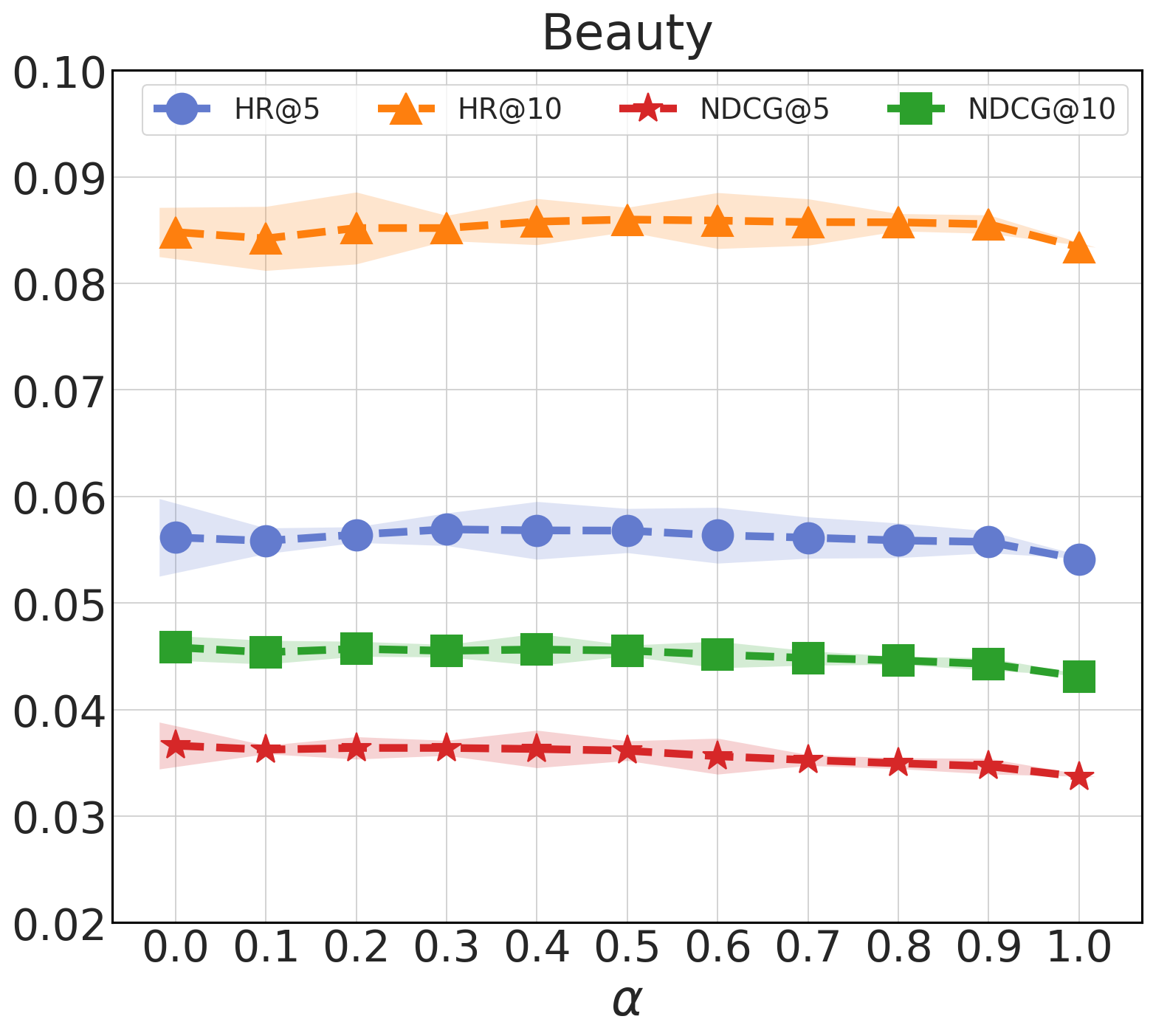}\label{fig:beauty_alpha}}
    \subfigure{
        \includegraphics[width=.309\linewidth]{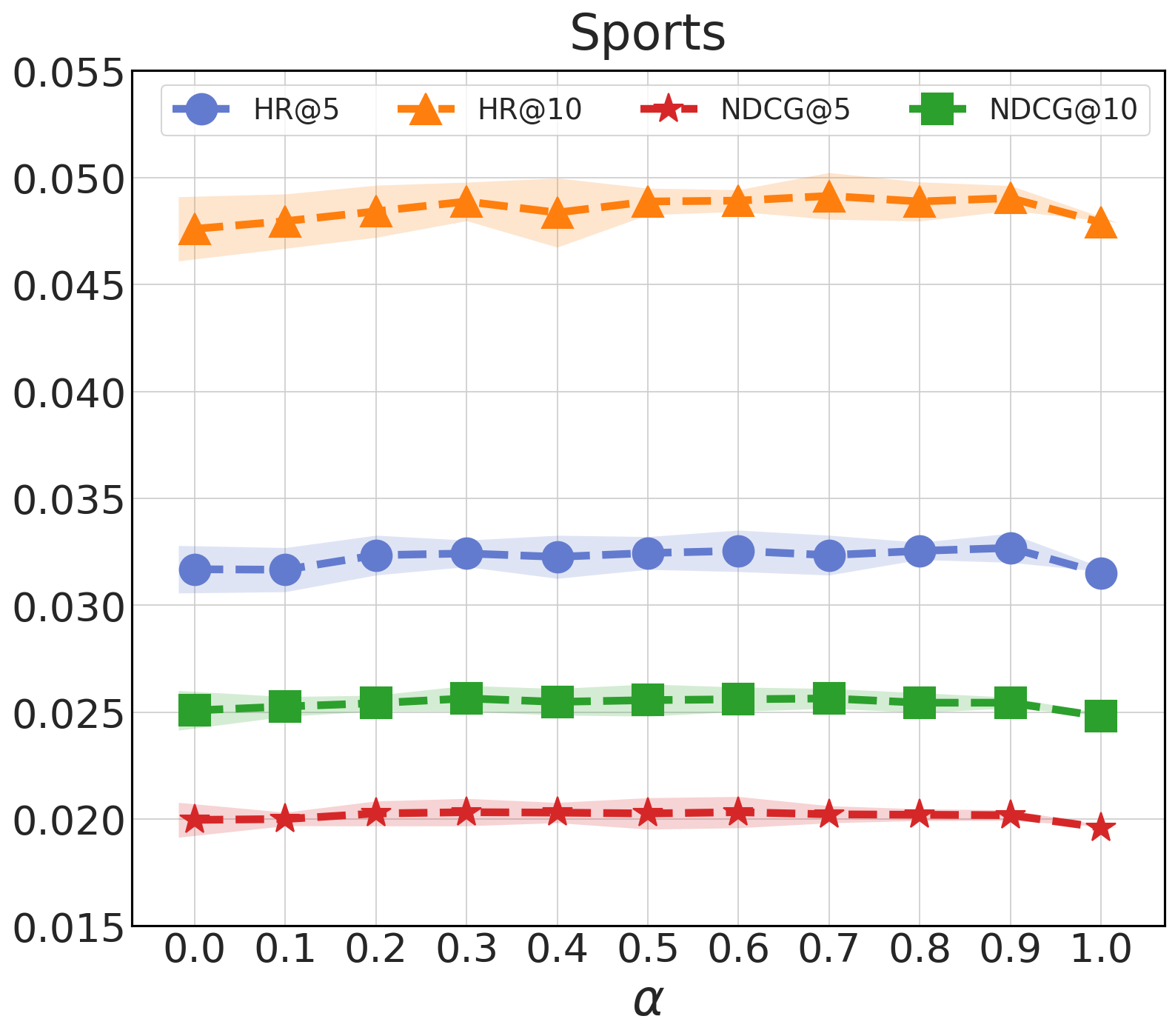}\label{fig:sports_alpha}}
     \subfigure{
        \includegraphics[width=.309\linewidth]{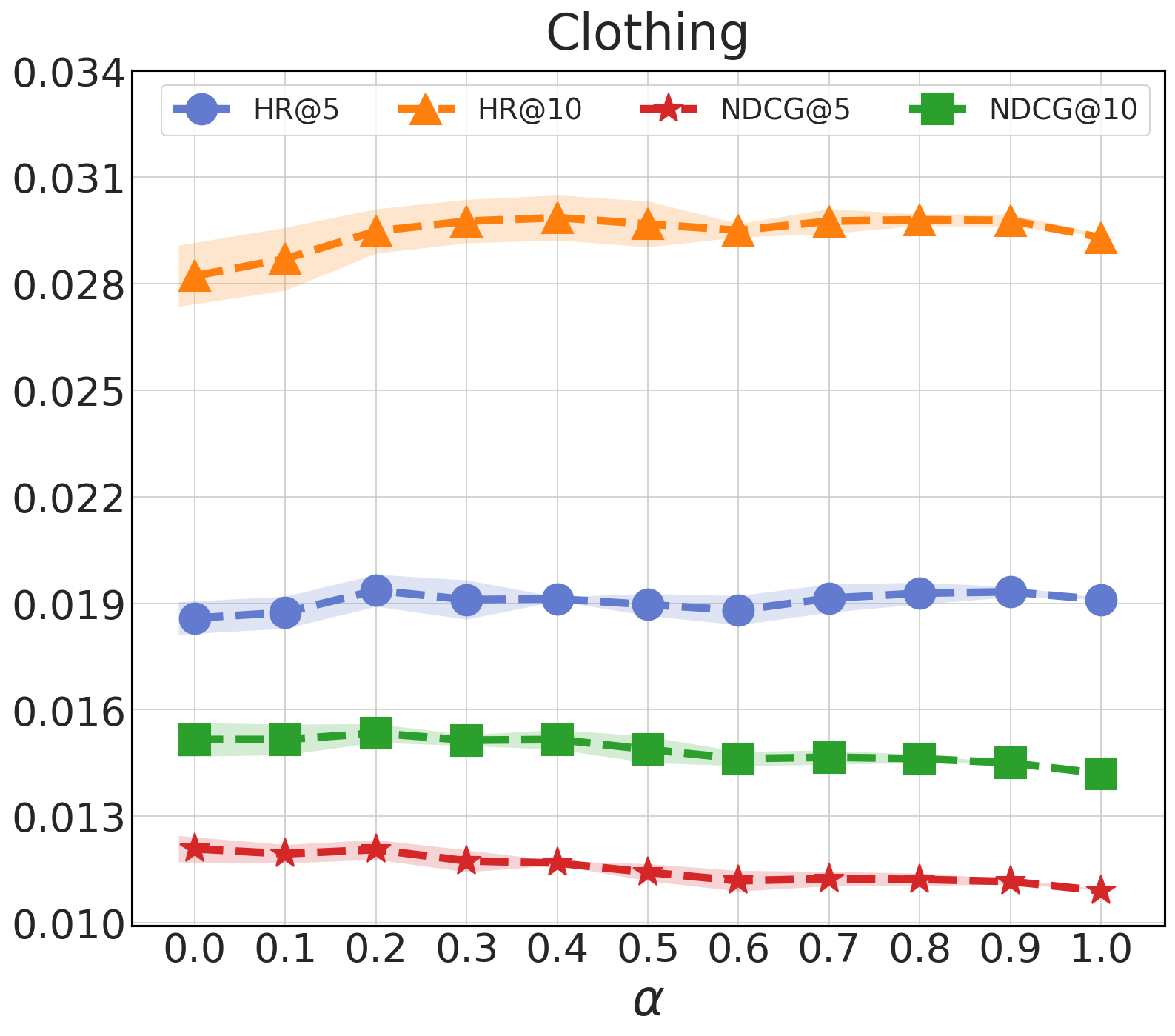}\label{fig:clothing_alpha}}

    \subfigure{
        \includegraphics[width=.309\linewidth]{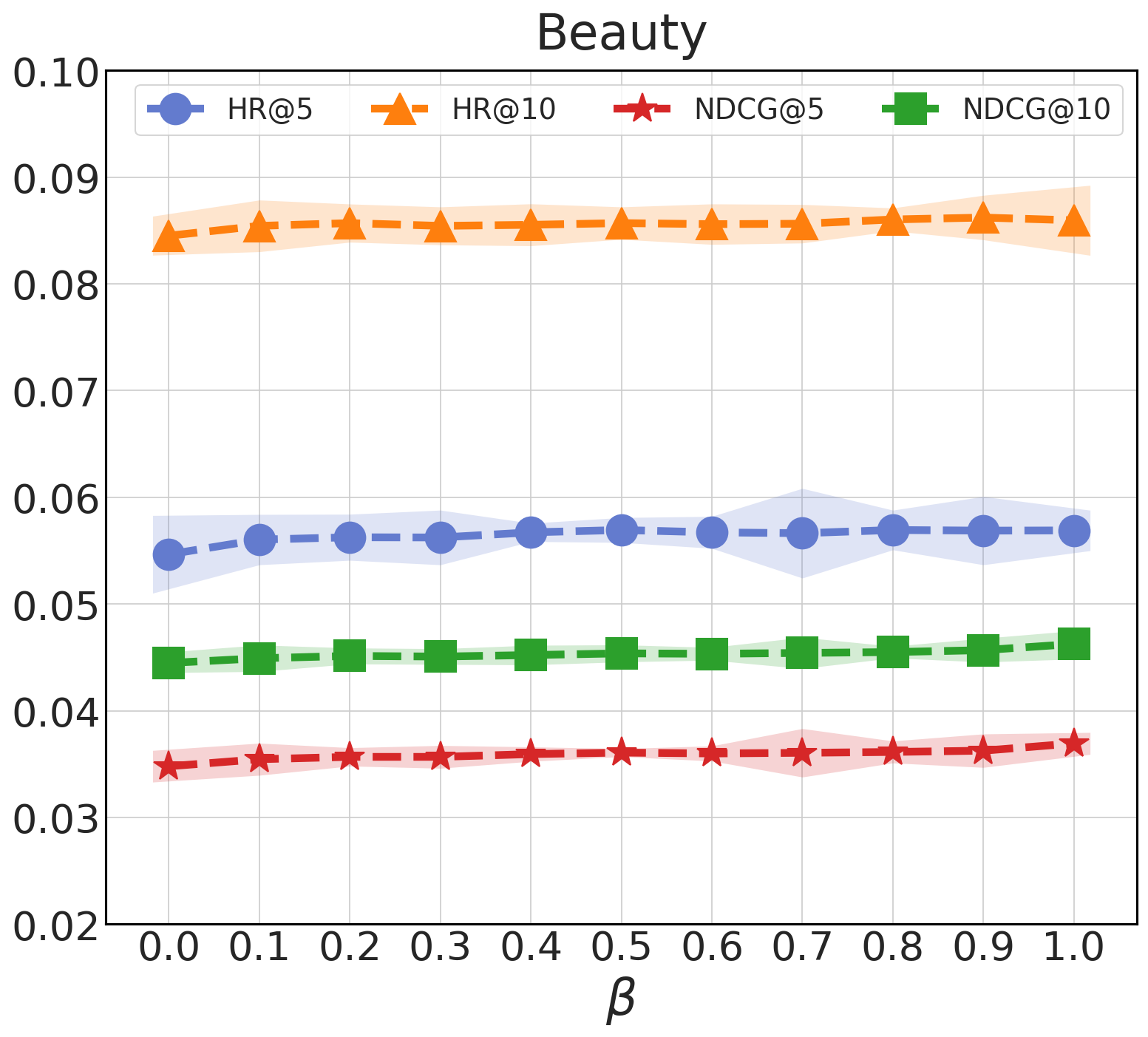}\label{fig:beauty_beta}}
    \subfigure{
        \includegraphics[width=.309\linewidth]{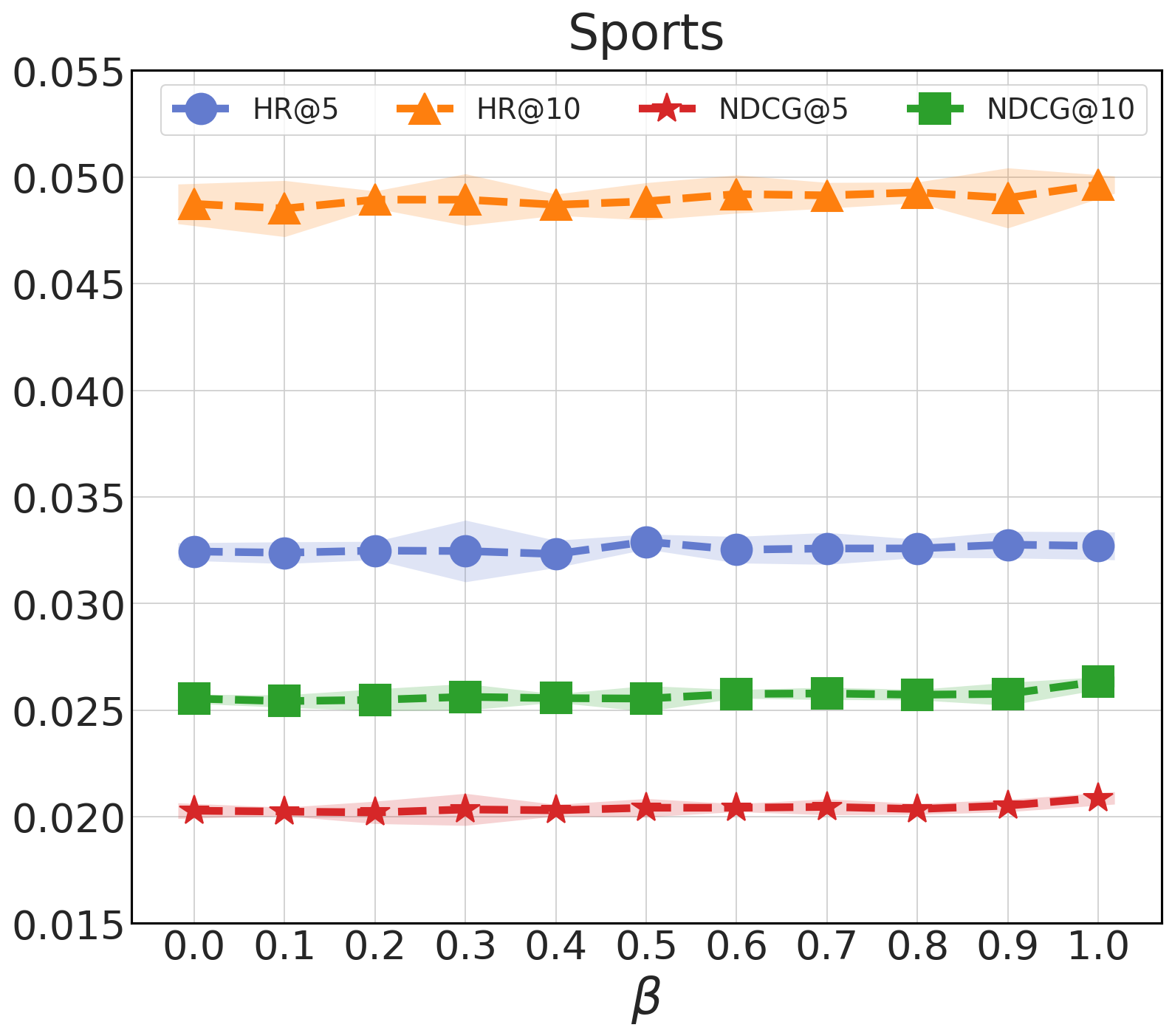}\label{fig:sports_beta}}
     \subfigure{
        \includegraphics[width=.309\linewidth]{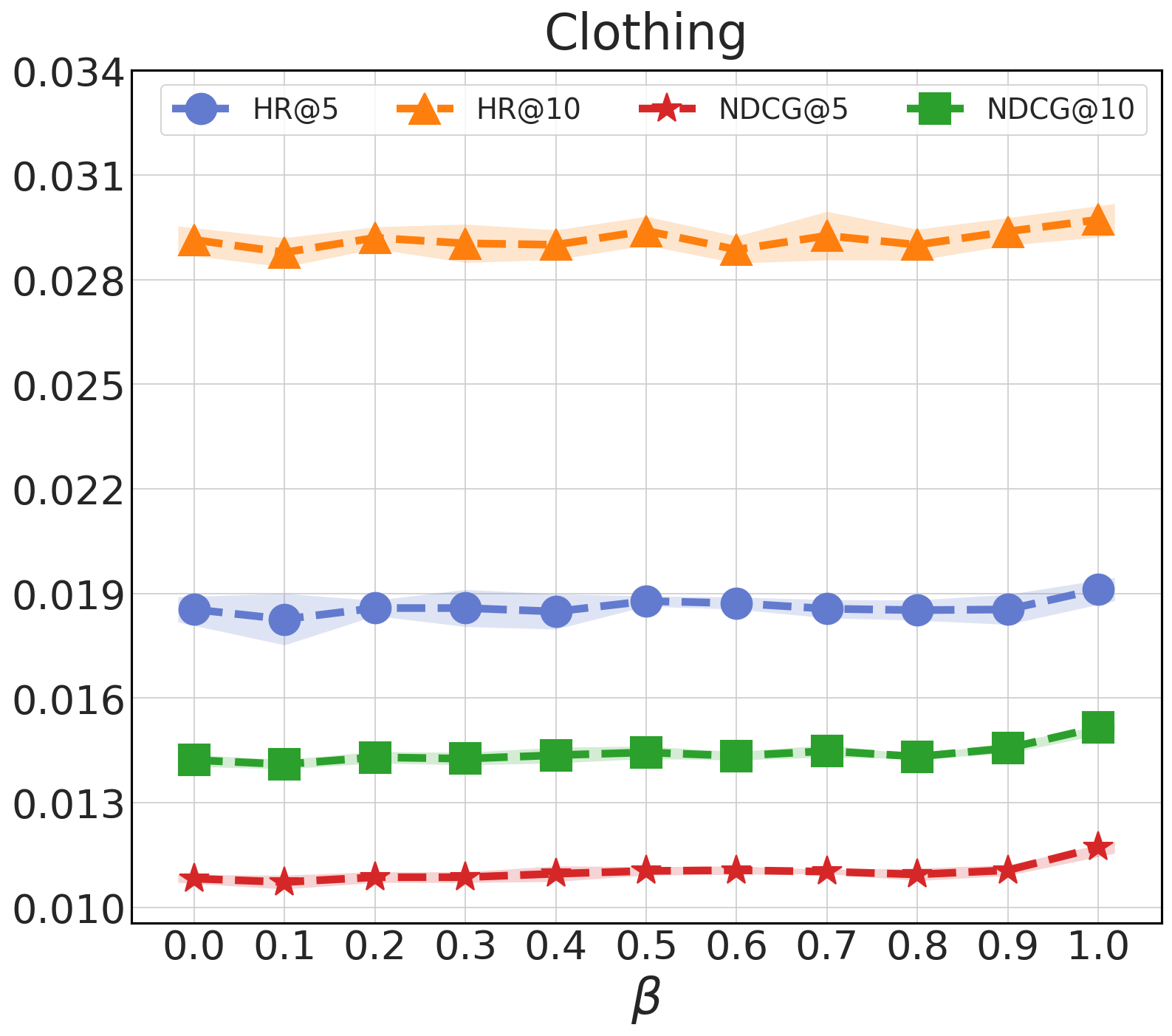}\label{fig:clothing_beta}}

    \subfigure{
        \includegraphics[width=.309\linewidth]{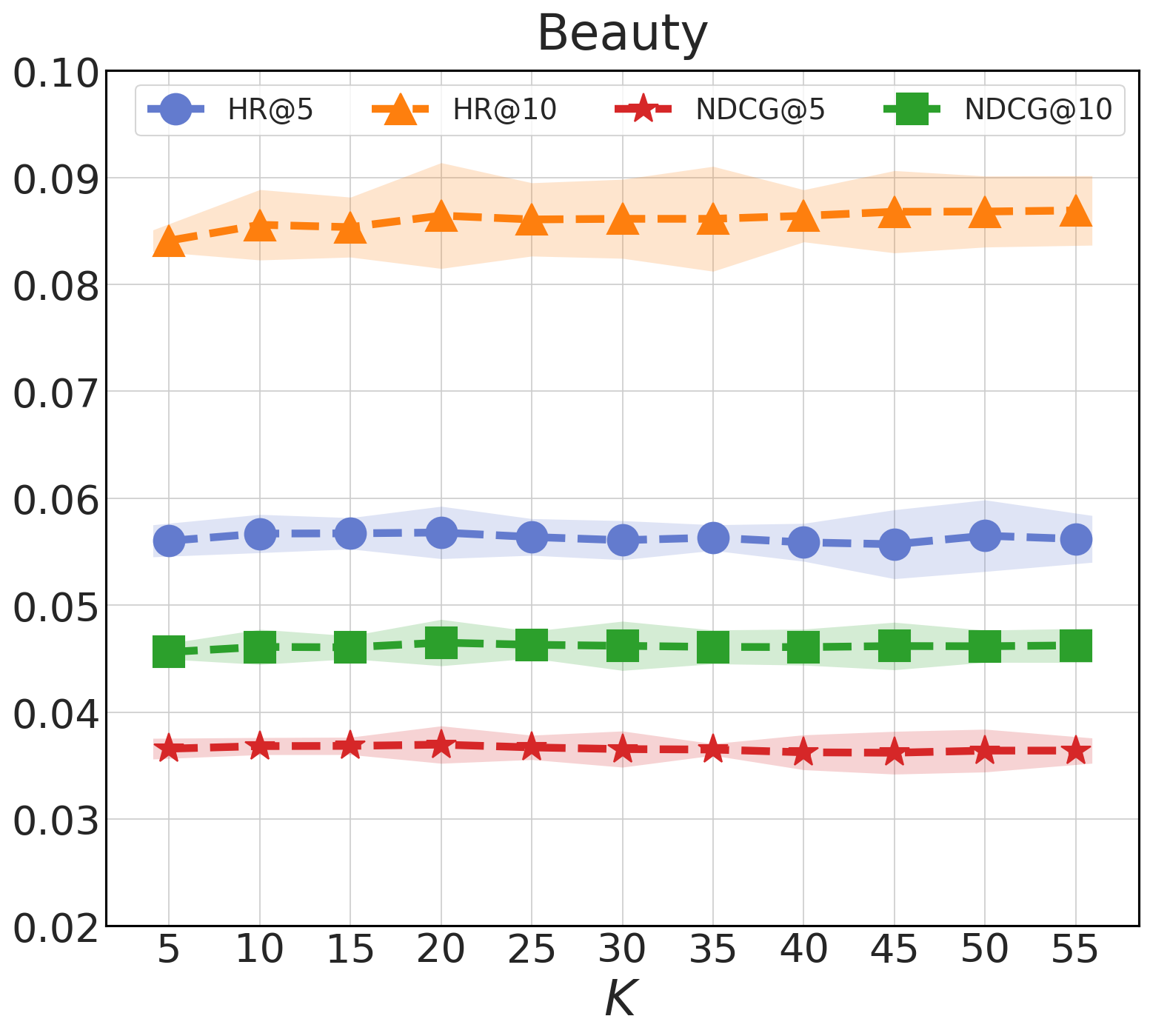}\label{fig:beauty_nk}}
    \subfigure{
        \includegraphics[width=.309\linewidth]{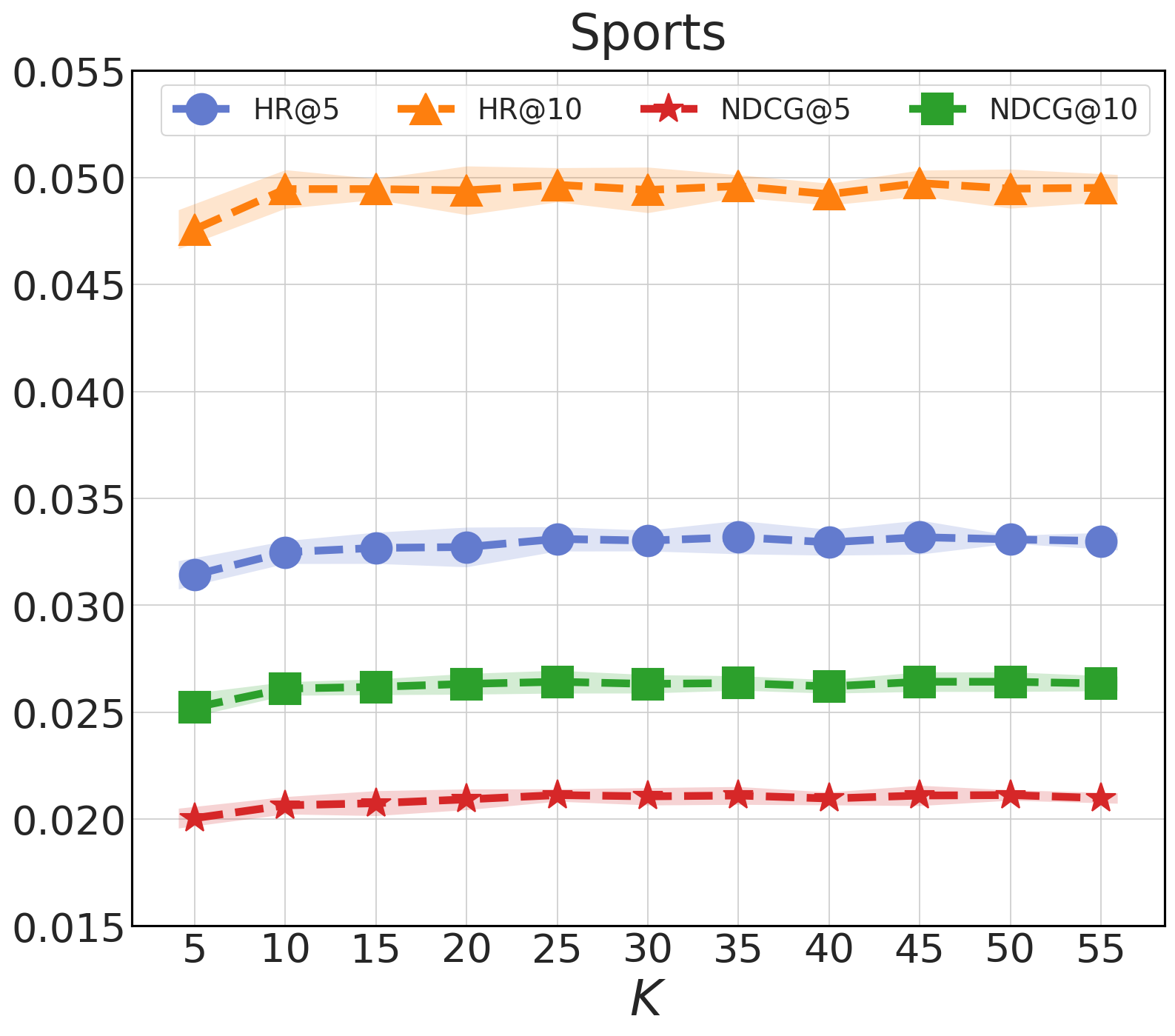}\label{fig:sports_nk}}
     \subfigure{
        \includegraphics[width=.309\linewidth]{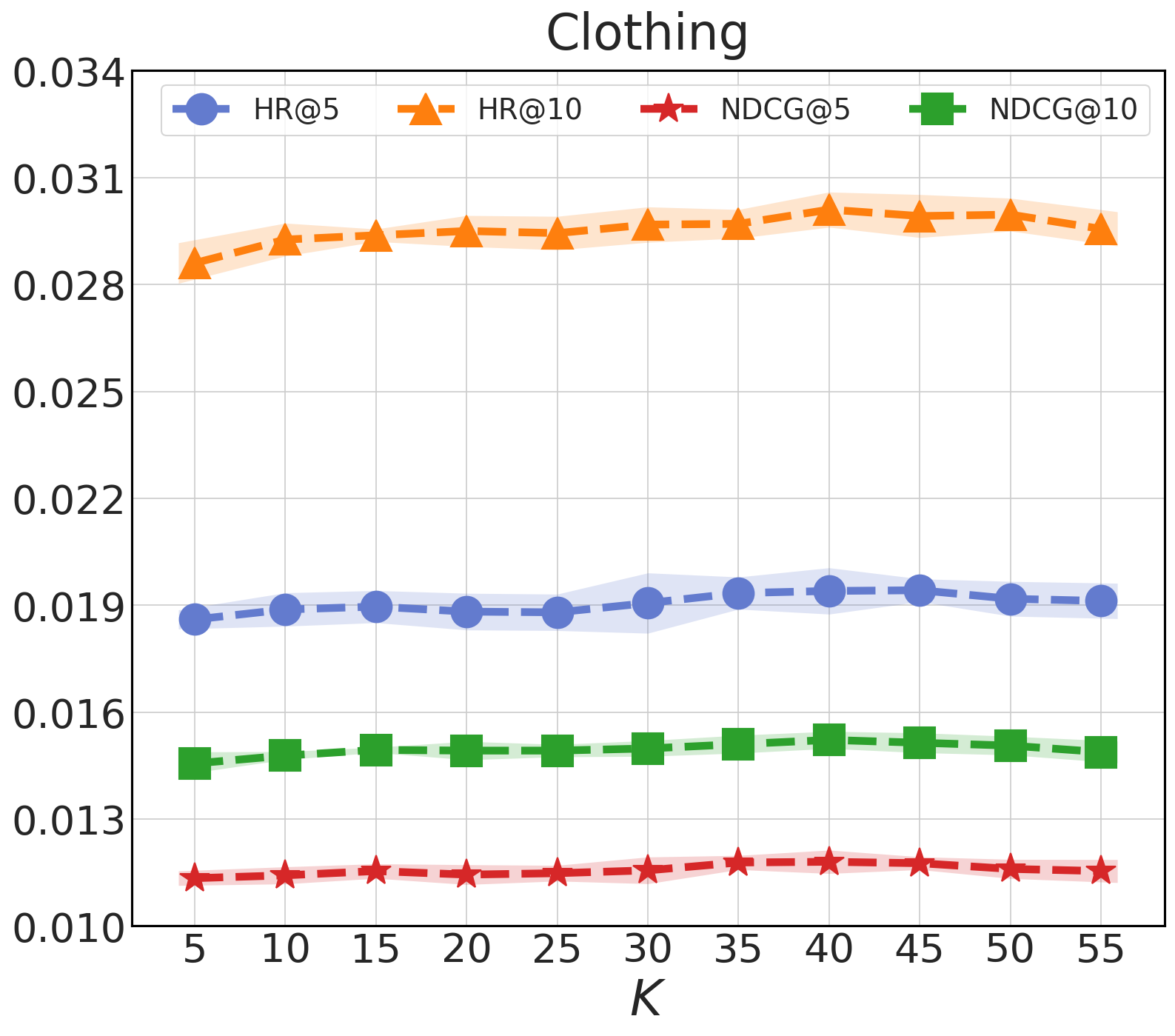}\label{fig:clothing_nk}}
    
    \caption{Parameter sensitivity \textit{w.r.t.} the control coefficients $\alpha, \beta$ and the number of retrieved memories $K$, where we plot the error bar with the mean $\pm$ $\gamma\times$standard deviation and $\gamma$ is set as $500\times$the scale interval of the y-axis.}
    \label{fig:parameter_alpha}
\end{figure*}
\begin{figure*}[t]
    \centering  
     \subfigure[Results on the Beauty dataset \textit{w.r.t.} HR@5 and NDCG@5.]{
        \includegraphics[width=0.99\linewidth]{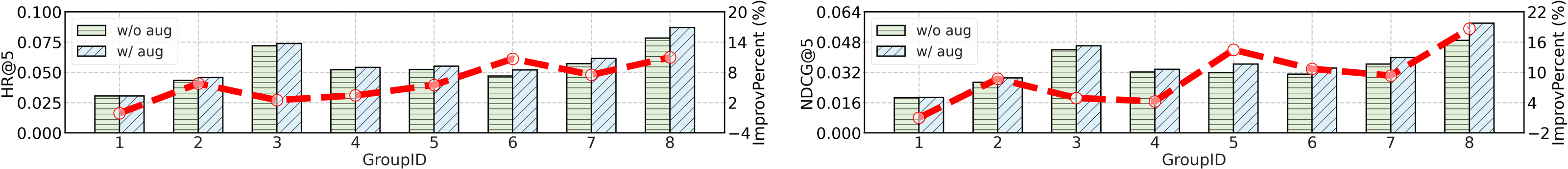}\label{fig:beauty_user_fren}}
    \subfigure[Results on the Sports dataset \textit{w.r.t.} HR@5 and NDCG@5.]{
        \includegraphics[width=0.99\linewidth]{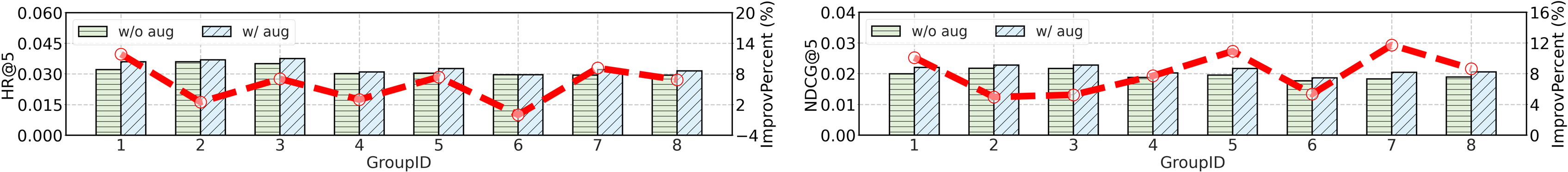}\label{fig:sports_user_fren}}
    \caption{Performance comparison over different user groups between the model with retrieval augmentation and without retrieval augmentation. The bar represents HR@5 and NDCG@5, while the line denotes the performance improvement percentage of ``w/ aug'' compared to ``w/o aug''.}
    \label{fig:user_fren}
\end{figure*}
\section{Parameter Sensitivity (RQ7)}
\label{app:para_sensi}
We study the effects of RaSeRec's key hyper-parameters on its performance, including the control coefficients $\alpha,\beta$ (Equ. (\ref{eq:fusion})) and the number of retrieved memories $K$ (\S \ref{sec:retri_ft}). For $\alpha$ and $\beta$, we vary them from 0 to 1.0, incrementing by 0.1; for $K$, we vary it from 5 to 55, incrementing by 5. 
We ran each experiment five times and computed the average as well as standard deviation results. The experimental results are shown in Figure \ref{fig:parameter_alpha}, where the line denotes the average results and the error bar denotes the scaled-up standard deviation results. From the results, we mainly have the following observations:
Firstly, we can observe that setting $\alpha$ to a too-big value (\textit{e.g.,} 1.0) or to a too-small value (\textit{e.g.,} 0.1) will significantly hurt the model performance. When alpha is high, implicit memories dominate the user representation, whereas if alpha is low, explicit memories dominate the user representation.
The above observation indicates the user representation should not be dominated by one side, \textit{i.e.,} implicit memories or explicit memories. In a nutshell, we suggest tuning $\alpha$ in the range of $0.3 \sim 0.7$. For example, setting $\alpha$ as 0.5 will lead to satisfactory performance on the Beauty dataset. 
Besides, we observe that the performance volatility of RaSeRec generally increases as $\alpha$ decreases. The main reason is that retrieved explicit memories are more diverse than implicit ones since the model backbone is frozen. In conclusion, we suggest setting $\alpha$ at around 0.5 to achieve stable recommendation performance and relatively diverse results.
Secondly, we can observe the results \textit{w.r.t.} different values of $\beta$ on all three datasets following similar trends. As $\beta$ increases, the recommendation performance and the performance volatility gradually increase. 
Generally, $\beta=0.9$ or $\beta=1.0$ leads to the best performance. It demonstrates that augmenting user representation via modeling the relationship between user representation and the retrieved user representation (Equ. (\ref{eq:first_channel})) can produce more informative and accurate user representation. 
We think the main reason is the model backbone learns to recommend items based on the user sequence instead of recommending users based on the target item. As such, users with similar behaviors lead to similar target items, whereas users who like the same item do not necessarily have similar behaviors. 
For the above reasons, augmenting user representation via modeling the relationship between user representation and the retrieved user representation (Equ. (\ref{eq:first_channel})) performs better than that between user representation and the corresponding target item embedding of the retrieved user representation (Equ. (\ref{eq:second_channel})). 
In conclusion, we suggest tuning $\beta$ in the range of $0.7 \sim 1.0$ carefully.

Thirdly, we observe that the performance of RaSeRec reaches the peak to different $K$ across three datasets, \textit{e.g.,} 20 is the best value for the Beauty dataset. 
This demonstrates the effectiveness of augmenting user representation via retrieved explicit memories and manifests that setting a suitable number of retrieved memories can considerably enhance performance. The results on the three datasets follow similar trends, where the performance gradually improves as $K$ increases and then starts to degrade when further increasing $K$. 
This phenomenon is also observed in long-context RAG \cite{jin2024long,DBLP:journals/corr/abs-2410-04343}, where the generative modeling performance initially increases and then declines when consistently increasing the number of retrieved passages.
The main reason is that increasing the number of retrieved memories will inevitably introduce irrelevant memories (\textit{i.e.,} ``noise'') that can mislead the retrieval-augmented module.
In conclusion, we suggest tuning $K$ in the range of $10 \sim 30$ for the Beauty and Sports datasets, and $30 \sim 50$ for the Clothing dataset. 
\section{Study on User Interaction Frequency (RQ8)}
\label{app:user_freq}
To investigate the robustness of RaSeRec against low-frequency (short user sequence) and high-frequency users (long user sequence), we split the test user sequences into 8 groups based on their length and ensure the number of test user sequences within each group is the same. The larger the GroupID, the longer the test user sequences. 
The experimental results are shown in Figure \ref{fig:user_fren}, where models with retrieval augmentation mention ``w/ aug'' otherwise ``w/o aug''. From the results, we mainly have the following observations:
Firstly, it is surprising that augmenting user representations with retrieved memories is highly effective for high-frequency users. For example, on the $8$-th group of the beauty dataset, the performance improvements of ``w/ aug'' over ``w/o aug'' are 10.96\% and 18.67\% in terms of HR@5 and NDCG@5, respectively. 
This demonstrates that even for high-frequency users, implicit memories may fall short of capturing user preferences accurately, while explicit memories can remedy this issue to a large extent.
Secondly, we find the tendency of improvements \textit{w.r.t.} different groups vary in datasets. For example, on the Beauty dataset, the improvement of high-frequency users is larger than that of low-frequency users. 
While on the sports dataset, different groups have similar performance improvements. We think the main reason is that the \#avg.length of the sports dataset is shorter than that of the beauty dataset. 
As such, there are a lot of short- and medium-term memories in the memory bank of the sports dataset, as shown in Table \ref{table:dataset_part}. These memories can largely facilitate the performance of low- and medium-frequency users.
Thirdly, we are surprised to find that there is no significant difference in the performance of each group on the sports dataset. However, on the beauty dataset, high-frequency users perform better than low-frequency ones, which is in accordance with people's intuition. We think the main reason is that the sports dataset is sparser than the beauty dataset. 
As such, there are not enough supervised signals for the model to learn high-quality user representation for long user sequences. This also explains why the model performs poorly when only the $\mathrm{L}$ partition is retained, as shown in Figure \ref{fig:sports_abla_memo}.
In a nutshell, we suggest performing retrieval augmentation for both low- and high-frequency users.

\end{document}